\documentclass[11pt]{article}
\usepackage{geometry}
\usepackage{color}
\usepackage{amsmath}
\usepackage{graphicx}
\usepackage{esint}
\usepackage[authoryear]{natbib}
\usepackage[unicode=true]
 {hyperref}

\makeatletter

\providecommand{\tabularnewline}{\\}

\@ifundefined{date}{}{\date{}}
\usepackage{amsmath}
\usepackage{amssymb}
\allowdisplaybreaks

\newcommand{\biblist}{\begin{list}{}
{\listparindent 0.0cm \leftmargin 0.50cm \itemindent -0.50 cm
\labelwidth 0 cm \labelsep 0.50 cm
\usecounter{list}}\clubpenalty4000\widowpenalty4000}
\newcommand{\ebiblist}{\end{list}}

\usepackage{latexsym}
\usepackage{multirow}

\newtheorem{lemma}{Lemma}\newtheorem{theorem}{Theorem}
\newtheorem{assumption}{Assumption}\newtheorem{remark}{Remark}

\newtheorem{example}{Example}
\newtheorem{definition}{Definition}[section]

\newcommand{\be}{\begin{equation}}
\newcommand{\en}{\end{equation}}
\newcommand{\bea}{\begin{eqnarray}}
\newcommand{\ena}{\end{eqnarray}}
\newcommand{\ba}{\begin{array}}

\newcommand{\ea}{\end{array}}

\newcommand{\R}{{\mathcal{R}}}

\newcommand{\pr}{\mathbb{P} }

\newcommand{\T}{\mathrm{\scriptscriptstyle T}}
 
\newcommand{\ATT}{ {\mathrm{ATT}}} 
\newcommand{\adj}{ {\mathrm{adj}}} 
\newcommand{\mat}{ {\mathrm{mat}}} 
\newcommand{\obs}{ {\mathrm{obs}}} 
 
\newcommand{\cov}{ {\mathrm{cov}}} 
\newcommand{\dsm}{ {\mathrm{dsm}}} 
\newcommand{\psm}{ {\mathrm{psm}}} 
\newcommand{\prog}{ {\mathrm{prog}}} 
\newcommand{\J}{ {\mathcal{J}}} 
 
\newcommand{\plim}{ {\mathrm{plim}}} 
\newcommand{\F}{ {\mathcal{F}}}

\newcommand{\N}{ {\mathcal{N}}} 
 
\newcommand{\It}{ \mathcal{I}} 
\newcommand{\logit}{ {\mathrm{logit}}} 
\newcommand{\de}{ {\mathrm{d}}} 
 
\newcommand{\dm}{ {d_{V}}} 
\newcommand{\E}{ {\mathbb{E}} } 
\newcommand{\bP}{ {\mathbb{P}}} 
\newcommand{\V}{ {\mathbb{V}}} 
\newcommand{\bone}{\mathbf{1}}
\newcommand{\cU}{ {\mathcal{U}}} 

\newcommand*{\indep}{%
  \mathbin{%
    \mathpalette{\@indep}{}%
  }%
}
\newcommand*{\nindep}{%
  \mathbin{
    \mathpalette{\@indep}{\not}
  }%
}
\newcommand*{\@indep}[2]{%
  \sbox0{$#1\perp\m@th$}
  \sbox2{$#1=$}
  \sbox4{$#1\vcenter{}$}
  \rlap{\copy0}
  \dimen@=\dimexpr\ht2-\ht4-.2pt\relax
  \kern\dimen@
  {#2}%
  \kern\dimen@
  \copy0 
}

\makeatother

\begin{document}
\baselineskip .3in
\title{\textbf{Multiply robust matching estimators of average and quantile
treatment effects}}
\author{Shu Yang\thanks{Department of Statistics, North Carolina State University, North Carolina
27695, U.S.A. Email: syang24@ncsu.edu}$\ \ $and Yunshu Zhang\thanks{Department of Statistics, North Carolina State University, North Carolina
27695, U.S.A. Email: yzhan234@ncsu.edu}}
\maketitle
\begin{abstract}
\textcolor{black}{}Propensity score matching has been a long-standing
tradition for handling confounding in causal inference, however requiring
stringent model assumptions. In this article, we propose double score
matching (DSM) for general causal estimands utilizing two balancing
scores including the propensity score and prognostic score. To gain
the protection of possible model misspecification, we posit multiple
candidate models for each score. We show that the de-biasing DSM estimator
achieves the multiple robustness property in that it is consistent
for the true causal estimand if any model of the propensity score
or prognostic score is correct.
\end{abstract}

\newpage{}

\section{Introduction}

Causal inference plays an important role in science, education, medicine,
policy, and economics. If all confounders of the treatment-outcome
relationship are observed, one can use standard techniques, such as
regression adjustment, inverse probability of treatment weighting
(IPW), augmented IPW (AIPW), and matching to adjust for confounding
\citep{imbens2015causal}. Among them, the AIPW estimator is most
popular because it achieves the so-called \textit{double robustness}
property by combining the use of models for the probability of treatment
assignment, also known as the propensity score \citep{rosenbaum1983central},
and the outcome mean function. More specifically, it consistently
estimates the treatment effect if either one of these functions is
modeled correctly \citep[e.g.,][]{lunceford2004stratification,bang2005doubly}.
However, inevitably, weighting estimators can have a high variability
by inverting the estimated propensity scores \citep[e.g.,][]{kang2007demystifying,guo2014propensity},
especially if these probabilities are close to zero or one. Matching
has multiple features that are more desirable than weighting:
\begin{enumerate}
\item matching does not involve weighting by the inverse of the propensity
score and therefore avoids the possibly large variability due to weighting
\citep{frolich2004finite};
\item matching is transparent and intuitively appealing with the goal of
replicating the ideal randomized experiment \citep{rosenbaum1989optimal,HeckmanIchimuraTodd1997,dehejia&wahba98,dehejia1999causal,rubin2006matched,stuart2010matching}; 
\item matching can be viewed as a hot deck imputation method which can provide
valid estimators of general parameters depending on the entire distribution,
such as quantiles \citep{ford1983overview}.
\end{enumerate}
Although matching has a substantial promise,\textcolor{black}{{} its
applications are still less popular compared to weighting, }partly
due to the issue of the curse of dimensionality. In the presence of
many covariates, matching directly on high-dimensional covariates
is incapable of removing all confounding biases. To overcome this
challenge, researchers have proposed different dimension reduction
techniques to facilitate matching. On the one hand, \citet{rosenbaum1983central}
demonstrated the central role of the propensity score as being a balancing
score in the sense that same propensity score distributions in different
treatment groups lead to same covariate distributions. Therefore,
matching solely on the propensity score (PSM) can remove all confounding
biases (e.g., \citealp{abadie2016matching}). On the other hand, \citet{hansen2008prognostic}
proposed an alternative balancing score: the prognostic score, also
called the disease risk score (i.e., a sufficient statistic for the
potential outcomes given which the potential outcomes and covariates
are independent). This score provides a balance of disease risks between
the treatment groups, as distinct from the balance of treatment propensities
provided by the propensity score. In economics, matching based on
prognostic score (PGM) has been previously proposed in \citet{imbens2004nonparametric}
and \citet{zhao2004using}, where the prognostic score is a vector
of linear predictors in treatment-specific outcome regressions. PGM
is also similar to predictive mean matching \citep{rubin1986statistical,yang2017predictive}
in the missing data literature to compensate for nonresponse. In the
comparative effectiveness research, PGM has been shown to be more
advantageous than PSM when the propensity score distributions are
strongly separated \citep{wyss2015matching,kumamaru2016dimension}.
\citet{smith2015time} extended PGM to a time-dependent treatment
setting. While PGM is gaining its popularity, it relies on a correctly
specification of the prognostic score \citep{wyss2015matching,wyss2017dry}.
As analogous to AIPW, it is advantageous to combine the use of the
propensity and prognostic score in matching \citep{hansen2008prognostic}.
\citet{leacy2014joint} showed empirically that the joint use of two
scores in matching (which we refer to as double score matching, DSM)
improves the treatment effect estimation. \citet{antonelli2018doubly}
later established the double robustness of matching jointly on propensity
and prognostic scores in the sense that the matching estimator is
consistent for the average treatment effect (ATE) if either one of
the score models is correctly specified. 

In this article, we propose new DSM estimators based on the propensity
score and the prognostic score. Because each score creates a balance
between the treated and control groups, the augmented score serves
as a ``double balancing score'' as shown by \citet{antonelli2018doubly}.
To estimate the ATEs, existing DSM would require adjusting for the
vector of the propensity score and possibly multiple treatment-specific
prognostic scores (i.e., one for each treatment group). Instead of
estimating the ATEs directly, we focus on estimating the average of
the potential outcomes separately for each treatment level, which
requires adjusting only for the propensity score and the prognostic
score for that particular level of the treatment. This insight allows
us to reduce the dimension of the double score further without giving
up the double balancing property. This strategy also plays an important
role for the dimension reduction for constructing improved DSM estimators.

In practice, the double score is unknown and therefore requires modeling
and estimation. Similar to \citet{antonelli2018doubly}, the new DSM
estimator is doubly robust, which includes one propensity score model
and one prognostic score model. With an unknown data generating process,
there is no guarantee that either of the two models is correctly specified.
To gain additional protection against model misspecification, we posit
multiple models for the propensity score and prognostic score. Doing
so, however, may introduce bias due to matching discrepancy based
on a moderately high-dimensional matching variable \textcolor{black}{\citep{abadie2011bias},}
although our strategy of estimating the average of potential outcomes
separately helps dimension reduction. In this case, we propose the
de-biasing DSM estimator that corrects for the bias due to matching
discrepancy. We show that the DSM estimator has the \textit{multiple
robustness} property, which guarantees the estimation consistency
if any one of the candidate models for the propensity score or prognostic
score is correctly specified. This result is similar in essence to
the multiply robust weighting estimators in the missing data and survey
literature \citep{han2013estimation,han2014multiply,chen2017multiply2,chen2017multiply}).

Because the double scores are estimated prior to matching, it is necessary
to account for the uncertainty due to parameter estimation. The theoretical
task is non-trivial. The typical Taylor expansion technique can not
be used, because of the non-smooth nature of matching. Our derivation
is based on the technique developed by \citet{andreou2012alternative},
which offers a general approach for deriving the limiting distribution
of statistics that involve estimated nuisance parameters. \textcolor{black}{This
technique has been successfully used in \citet{abadie2016matching}
for the matching estimators based on the estimated propensity score.}
We extend their results to the DSM estimators requiring any one of
the models to be correctly specified.

The current matching literature has focused primarily on estimating
the ATEs; however, other aspects of the distribution such as quantiles
may be more appropriate in certain applications. For example, a treatment
strategy may not decrease average health cost but instead lowers the
upper tail of the cost distribution, so focusing only on ATEs would
not reveal the beneficial effect of the treatment strategy. In these
cases, it is more informative to study quantile treatment effects
(QTEs), which are defined as the differences in population quantiles
of the potential outcome distributions. Taking the advantage of matching
as a hot deck imputation method, we extend the multiply robust DSM
framework to estimate QTEs. 

The rest of this paper proceeds as follows. Section \ref{sec:Basic-Setup}
introduces notation, assumptions, and lemmas for various balancing
scores. Section \ref{sec:Double-score-matching} provides the new
perspective of using the double score as a dimension reduction tool
and proposes the new DSM estimator of the ATE with multiple candidate
models for the double score. Section \ref{sec:Main-results} establishes
the multiple robustness of the DMS estimator and its limiting distribution,
which allows quantifying the impact of the nuisance parameter estimation.
Section~\ref{sec:QTE} extends the DSM framework to the estimation
of the QTE. Section \ref{sec:Simulation-study} uses simulation to
evaluate the finite-sample properties of the DMS estimators. The simulation
results demonstrate that matching estimators outperform weighting
estimators. Section \ref{sec:Real-data-application} applies the DMS
estimators to an observational study from the job training program.
Section \ref{sec:Discussion} concludes, the supplementary material
contains the proofs and extensions to the average treatment effect
on the treated (ATT) and quantile treatment effect on the treated
(QTT), and an R package \textit{dsmatch} is available at \href{https://github.com/Yunshu7/dsmatch}{https://github.com/Yunshu7/dsmatch}.

\section{Notation, assumptions, and balancing scores \label{sec:Basic-Setup}}

Let $X_{i}$ be a vector of pre-treatment covariates, $A_{i}$ the
binary treatment, and $Y_{i}$ the outcome for unit $i=1,\ldots,n$.
We follow the potential outcomes framework. Let $Y_{i}(a)$ be the
potential outcome had unit $i$ been given treatment $a$ ($a=0,1$).
The observed outcome is $Y_{i}=Y_{i}(A_{i})=A_{i}Y_{i}(1)+(1-A_{i})Y_{i}(0)$.
We assume that $\{X_{i},A_{i},Y_{i}(0),Y_{i}(1)\}$, $i=1,\ldots,n$,
are independent and identically distributed. Thus, $(X_{i},A_{i},Y_{i})$,
$i=1,\ldots,n$, are also independent and identically distributed. 

Various causal estimands are useful to provide a comprehensive assessment
of treatment effects. The ATE is $\tau=\E\{Y(1)-Y(0)\}$. For $\xi\in(0,1),$
the overall $\xi$-QTE is $\Delta_{\xi}=q_{1,\xi}-q_{0,\xi}$, where
$q_{a,\xi}=\inf_{q}[\bP\{Y(a)\leq q\}\geq\xi]$, $a=0,1$. When the
outcome data follow a skewed distribution, QTEs may be more informative
measures of treatment effect. Similarly, the ATT is $\tau_{\mathrm{ATT}}=\E\{Y(1)-Y(0)\mid A=1\}$,
and the QTT is $\Delta_{\mathrm{QTT},\xi}=q_{1,\xi\mid A=1}-q_{0,\xi\mid A=1}$,
where $q_{a,\xi\mid A=1}=\inf_{q}[\bP\{Y(a)\leq q\}\geq\xi\mid A=1]$,
$a=0,1$. We focus on estimating $\tau$ and $\Delta_{\xi}$ in the
main text and provide extensions to the ATT and QTT in the supplementary
material. 

For simplicity of exposition, for a generic variable $V$, denote
\[
\mu_{a}(V)=\E\{Y(a)\mid V\},\ \sigma_{a}^{2}(V)=\V\{Y(a)\mid V\},\ e(V)=\bP(A=1\mid V),
\]
where $\mu_{a}(V)$ is an outcome mean function, $\sigma_{a}^{2}(V)$
is a variance function, and $e(V)$ is the propensity score. 

We focus on the setting where the standard positivity and treatment
ignorability assumptions hold \citep{rosenbaum1983central}.

\begin{assumption}\label{asump-overlap}There exist constants $c_{1}$
and $c_{2}$ such that $0<c_{1}\leq e(X)\leq c_{2}<1$ almost surely.

\end{assumption}

Assumption \ref{asump-overlap} implies a sufficient overlap of the
covariate distribution between the treatment groups. If this assumption
is violated, a common approach is to trim the sample; see \citet{yang2018trimming}.

\begin{assumption}\label{asump-ignorable} $\{Y(0),Y(1)\}\indep A\mid X$,
where $\indep$ means ``independent of''. \end{assumption}

Assumption \ref{asump-ignorable} has no testable implications. Generally,
it can be made more plausible by collecting detailed information on
characteristics of the units that are related to treatment assignment
and outcome. As a result, the dimension of $X$ may be high.

Under Assumptions \ref{asump-overlap} and \ref{asump-ignorable},
\[
\tau=\E\{\E(Y\mid A=1,X)-\E(Y\mid A=0,X)\},
\]
are identifiable from the joint distribution of the observed data
$(A,X,Y)$.

The seminal paper \citep{rosenbaum1983central} showed the central
role of the propensity score as a balancing score. 

\begin{lemma}[Propensity score as a balancing score; Rosenbaum and Rubin, 1983]\label{lem1:ps}Under
Assumptions \ref{asump-overlap} and \ref{asump-ignorable}, $\{Y(0),Y(1)\}\indep A\mid e(X)$.

\end{lemma}

Lemma \ref{lem1:ps} implies that 
\begin{equation}
\tau=\E[\E\{Y\mid A=1,e(X)\}-\E\{Y\mid A=0,e(X)\}].\label{eq:ps-balance}
\end{equation}
Then we can estimate $\tau$ through PSM, subclassification or weighting
\citep[e.g.,][]{imbens2015causal}.

On the other hand, \citet{hansen2008prognostic} introduced the notion
of the prognostic score.

\begin{definition}[Prognostic score;  Hansen, 2008]\label{def:progs}

The prognostic score $\Psi_{a}(X)$ is a sufficient statistic for
$Y(a)$ in the sense that $Y(a)\indep X\mid\Psi_{a}(X)$ for $a=0,1$.

\end{definition}

We illustrate the prognostic score in the following examples. 

\begin{example}\label{eg:glm}If $Y(a)$ follows a generalized linear
model with mean $\mu_{a}(X)=X^{\T}\beta_{a}$, then $\Psi_{a}(X)=X^{\T}\beta_{a}$
for $a=0,1$. 

\end{example}

\begin{example}\label{eg:location-shifted}If $Y(a)$ follows a location-shift
family $f_{a}\{y-\mu_{a}(X)\}$, then $\Psi_{a}(X)=\mu_{a}(X)$ for
$a=0,1$. 

\end{example}

\citet{hansen2008prognostic} showed that the prognostic score serves
an alternative balancing score. 

\begin{lemma}[Prognostic score as a balancing score; Hansen, 2008]\label{lem1:pgm}Under
Assumptions \ref{asump-overlap} and \ref{asump-ignorable}, $\{Y(0),$
$Y(1)\}\indep A\mid\Psi(X)$, where $\Psi(X)=\{\Psi_{0}(X),\Psi_{1}(X)\}$.

\end{lemma}

Lemma \ref{lem1:pgm} implies that 
\begin{equation}
\tau=\E[\E\{Y\mid A=1,\Psi(X)\}-\E\{Y\mid A=0,\Psi(X)\}].\label{eq:prog-balance}
\end{equation}
Then we can estimate $\tau$ through PGM and subclassification to
remove the confounding biases. 

Combining the propensity score and the prognostic score, \citet{antonelli2018doubly}
showed that the double score is also a balancing score in the sense
that treatment ignorability maintains by conditioning on the double
score. 

\begin{lemma}[Double score as a balancing score; Antonelli et al, 2018]\label{lem2:ds}Under
Assumptions \ref{asump-overlap} and \ref{asump-ignorable}, $\{Y(0),$
$Y(1)\}\indep A\mid\{h(X),\Psi(X)\}$ and $\{Y(0),$ $Y(1)\}\indep A\mid\{e(X),h(X)\}$
for any $h(X)$.

\end{lemma}

Lemma \ref{lem2:ds} implies that
\begin{eqnarray}
\tau & = & \E[\E\{Y\mid A=1,e(X),\Psi(X)\}-\E\{Y\mid A=0,e(X),\Psi(X)\}].\label{eq:ds-bal}
\end{eqnarray}
Then we can estimate $\tau$ through DSM or subclassification based
on the double score. 

DSM is an attractive alternative to PSM; however, the dimension reduction
property of Lemma \ref{lem2:ds} depends on the dimension of the prognostic
score. In Example \ref{eg:glm}, the dimension of the prognostic score
is two; thus, the dimension of the double score is three. \citet{antonelli2018doubly}
made an additional assumption that there does not exist treatment
effect modification, under which the prognostic score is a scalar.
Nonetheless, such additional assumptions may be controversial. The
problem is that if the dimension of the matching variable is higher,
the bias order of the matching estimator becomes larger, see Section
\ref{subsec:General-matching-estimators}, suggesting that the advantages
of PSM do not carry over to DSM. To preserve the simplicity of matching
(avoiding de-biasing), we show that further improvement of the dimension
reduction property of the double score is possible without additional
assumptions. Then, the advantage of PSM carries over to DSM, see Section
\ref{subsec:New-simple-double}. Moreover, because the double score
is unknown in practice, one must posit models and estimate the double
score from the observed data. To gain robustness to model misspecification,
we posit multiple candidate models for the double score. We propose
a multiply robust DSM procedure and show that the de-biasing matching
estimator achieves multiple robustness, see Section \ref{subsec:New-multiply-robust}. 

\section{DSM estimators of the ATE\label{sec:Double-score-matching}}

\subsection{General matching estimators\label{subsec:General-matching-estimators}}

To fix ideas, we consider matching with replacement with the number
of matches fixed at $M$. Matching estimators hinge on imputing the
missing potential outcome for each unit. In practice, the most common
choice of $M$ is $1$, then the matching procedure becomes nearest
neighbor imputation \citep{little2002statistical,chen2000nearest,chen2001jackknife}.
To be precise, for unit $i$, the potential outcome under $A_{i}$
is the observed outcome $Y_{i};$ the (counterfactual) potential outcome
under $1-A_{i}$ is not observed but can be imputed by the observed
outcomes of the nearest $M$ units with $1-A_{i}$. 

To illustrate the properties of the matching estimator, we first consider
a generic variable $V$ as the matching variable. Table \ref{tab:Matching-schemes}
summarizes the choices of $V$. To stabilize the numerical performance,
it is desirable to standardize $V$ such that each component has mean
zero and variance one. Without loss of generality, we use the Euclidean
distance to determine neighbors; the discussion applies to other distances
\citep{abadie2006large}. \textcolor{black}{We denote $\J_{V,i}$
as the index set for these matched units for unit $i$} and $K_{V,i}=\sum_{l=1}^{n}\bone(i\in\J_{V,l})$
as the number of times that unit $i$ is used as a match, where the
subscript ``$V$'' in $\J_{V,i}$ and $K_{V,i}$ indicates the name
of the matching variable. Table \ref{tab:Matching-schemes} (I) illustrates
the above matching scheme to impute the missing potential outcomes.
For unit $i$ with $A_{i}=1$, the imputed potential outcomes are
$\hat{Y}_{i}(1)=Y_{i}$ and $\hat{Y}_{i}(0)=M^{-1}\sum_{j\in\J_{V,i}}Y_{j}$.
For unit $i'$ with $A_{i'}=0$, $\hat{Y}_{i'}(1)=M^{-1}\sum_{j\in\J_{V,i'}}Y_{j}$
and $\hat{Y}_{i'}(0)=Y_{i'}$. Once we approximate both potential
outcomes for all units, a simple matching estimator of $\tau$ is
\[
\hat{\tau}_{\mat}=n^{-1}\sum_{i=1}^{n}\{\hat{Y}_{i}(1)-\hat{Y}_{i}(0)\}=n^{-1}\sum_{i=1}^{n}(2A_{i}-1)(1+M^{-1}K_{V,i})Y_{i}.
\]
To establish the asymptotic properties of $\hat{\tau}_{\mat}$, \citet{abadie2006large}
derived the following decomposition
\[
n^{1/2}(\hat{\tau}_{\mat}-\tau)=B_{n}+C_{n},
\]
where
\begin{eqnarray}
B_{n} & = & n^{-1/2}\sum_{i=1}^{n}(1-2A_{i})\left[M^{-1}\sum_{j\in\J_{V,i}}\left\{ \mu_{A_{i}}(V_{i})-\mu_{A_{i}}(V_{j})\right\} \right],\label{eq:B2}\\
C_{n} & = & n^{-1/2}\sum_{i=1}^{n}\left[\mu_{1}(V_{i})-\mu_{0}(V_{i})-\tau+(2A_{i}-1)\left(1+M^{-1}K_{V,i}\right)\left\{ Y_{i}-\mu_{A_{i}}(V_{i})\right\} \right].\nonumber 
\end{eqnarray}
By Assumptions \ref{asump-overlap} and \ref{asump-ignorable}, for
$V$ to be $X$, propensity score, prognostic score or double score,
we have $\E\{\mu_{1}(V)-\mu_{0}(V)\}=\tau$, and therefore $\E(C_{n})=0$.
The difference $\mu_{A_{i}}(V_{i})-\mu_{A_{i}}(V_{j})$ in (\ref{eq:B2})
accounts for the matching discrepancy, so $B_{n}$ contributes to
the asymptotic bias of the matching estimator. In general, if the
matching variable is $\dm$-dimensional, we have $\E(B_{n})=O(n^{1/2-2/\dm})$
(\citealp{abadie2006large}, Theorem 1). Table \ref{tab:bias order}
demonstrates the relationship of the bias order and $\dm$. If $\dm\geq4$,
the bias is non-negligible. If $\dm=3$, the bias shrinks to zero
as $n$ increases but the convergence rate $-1/6$ is slow. If $\dm=2$
and $1$, the bias shrinks to zero at much faster rates $-1/2$ and
$-3/2$, respectively. Therefore, in finite samples, matching based
on a $3$-dimensional double score $\{e(X),\Psi(X)\}$ is likely to
have a noticeable bias. Reducing $\dm$ to $2$ or $1$ is worthwhile
to make the bias achieve faster rates of converging to zero. 

\begin{table}[h]
\begin{centering}
{\scriptsize{}\caption{\label{tab:Matching-schemes}Two matching schemes for imputing potential
outcomes. \textcolor{black}{$\J_{V,i}$ denotes the index set for
the matched units for unit $i$, where the subscript ``$V$'' represents
the name of the matching variable.} In (I), the matching variable
$V$ is the same for imputing the missing values of $Y(0)$ and $Y(1)$.
In (II), the matching variables $V_{0}$ and $V_{1}$ are different
for imputing the missing values of $Y(0)$ and $Y(1).$\textcolor{black}{{} }}
}{\scriptsize\par}
\par\end{centering}
\vspace{0.25cm}

\centering

\resizebox{\textwidth}{!}{
\begin{centering}
\begin{tabular}{lcccccccccc}
\hline 
\multicolumn{5}{c}{(I) Matching imputation} &  & \multicolumn{5}{c}{(II) New matching imputation}\tabularnewline
\cline{1-5} \cline{2-5} \cline{3-5} \cline{4-5} \cline{5-5} \cline{7-11} \cline{8-11} \cline{9-11} \cline{10-11} \cline{11-11} 
Unit & $A$ & $Y$ & $\hat{Y}(0)$ & $\hat{Y}(1)$ &  & Unit & $A$ & $Y$ & $\hat{Y}(0)$ & $\hat{Y}(1)$\tabularnewline
\cline{1-5} \cline{2-5} \cline{3-5} \cline{4-5} \cline{5-5} \cline{7-11} \cline{8-11} \cline{9-11} \cline{10-11} \cline{11-11} 
$i$ & $0$ & $Y_{i}$ & $Y_{i}$ & $M^{-1}\sum_{l\in\J_{\mathbf{V},i}}Y_{l}$ &  & $i$ & $0$ & $Y_{i}$ & $Y_{i}$ & $M^{-1}\sum_{l\in\J_{\mathbf{V_{1}},i}}Y_{l}$\tabularnewline
$i'$ & $1$ & $Y_{i'}$ & $M^{-1}\sum_{l\in\J_{\mathbf{V},i'}}Y_{l}$ & $Y_{i'}$ &  & $i'$ & $1$ & $Y_{i'}$ & $M^{-1}\sum_{l\in\J_{\mathbf{V_{0}},i'}}Y_{l}$ & $Y_{i'}$\tabularnewline
\hline 
 &  &  &  &  &  &  &  &  &  & \tabularnewline
 &  &  &  &  &  &  &  &  &  & \tabularnewline
\end{tabular}
\par\end{centering}
}

\resizebox{\textwidth}{!}{

\begin{tabular}{lcccccc}
\hline 
 & \multicolumn{2}{c}{(I) Matching Variable} &  & \multicolumn{3}{c}{(II) Matching Variable}\tabularnewline
\cline{2-3} \cline{3-3} \cline{5-7} \cline{6-7} \cline{7-7} 
 & $V$ & $\dm$ &  & $V_{0}$ & $V_{1}$ & $\dm$\tabularnewline
\cline{2-3} \cline{3-3} \cline{5-7} \cline{6-7} \cline{7-7} 
M.X & $X$ & $\dim(X)$ &  & $X$ & $X$ & $\dim(X)$\tabularnewline
PSM & $e(X)$ & $1$ &  & $e(X)$ & $e(X)$ & $1$\tabularnewline
PGM & $\{\Psi_{0}(X),\Psi_{1}(X)\}$ & $2$ &  & $\Psi_{0}(X)$ & $\Psi_{1}(X)$ & $1$\tabularnewline
DSM & $S=\{e(X),\Psi_{0}(X),\Psi_{1}(X)\}$ & $3$ &  & $S_{0}=\{e(X),\Psi_{0}(X)\}$ & $S_{1}=\{e(X),\Psi_{1}(X)\}$ & $2$\tabularnewline
DSM & $S=\{e^{j}(X),\Psi_{0}^{k}(X),\Psi_{1}^{k}(X):$ & $J+2K$ &  & $S_{0}=\{e^{j}(X),\Psi_{0}^{k}(X):$ & $S_{1}=\{e^{j}(X),\Psi_{1}^{k}(X):$ & $J+K$\tabularnewline
 & $j=1,\ldots J;k=1,\ldots,K\}$ &  &  & $j=1,\ldots J;k=1,\ldots,K\}$ & $j=1,\ldots J;k=1,\ldots,K\}$ & \tabularnewline
\hline 
\end{tabular}

}
\end{table}

\begin{table}[h]
\begin{centering}
{\scriptsize{}\caption{\label{tab:bias order}The order of bias of the matching variable
in terms of the dimension of the matching variable}
}{\scriptsize\par}
\par\end{centering}
\centering{}
\centering{}%
\begin{tabular}{lccccc}
\hline 
$\dm$ & $1$ & $2$ & $3$ & $4$ & $>4$\tabularnewline
\hline 
$O(n^{1/2-2/\dm})$ & $O(n^{-3/2})$ & $O(n^{-1/2})$ & $O(n^{-1/6})$ & $O(1)$ & $O(n^{1/10})$\tabularnewline
\hline 
\end{tabular}
\end{table}

\subsection{New simple DSM estimator\label{subsec:New-simple-double}}

Lemma \ref{lem3:newds} is the simple but key result. 

\begin{lemma}\label{lem3:newds}

Under Assumptions \ref{asump-overlap} and \ref{asump-ignorable},
$Y(a)\indep A\mid\{h(X),\Psi_{a}(X)\}$, $Y(a)\indep A\mid\{e(X),h(X)\}$
for any $h(X)$ and $a=0,1$.

\end{lemma}

Lemma \ref{lem3:newds} implies that
\begin{equation}
\E\{Y(a)\}=\E\left[\E\{Y(a)\mid e(X),\Psi_{a}(X)\}\right]=\E\left[\E\{Y\mid A=a,e(X),\Psi_{a}(X)\}\right],\ \ (a=0,1).\label{eq:newds-balancing}
\end{equation}

For its interpretation, it is useful to compare it to the result (\ref{eq:ds-bal})
by Lemma \ref{lem2:ds}. By Lemma \ref{lem2:ds}, we create subpopulations
where we can simultaneously compare the treated units and the control
units, leading to (\ref{eq:ds-bal}). These subpopulations were defined
by common values for $\{e(X),\Psi(X)\}$. By Lemma \ref{lem3:newds},
we do not construct such populations. The key insight is that in order
to estimate $\tau,$ it is not necessary to do so. Instead, we construct
subpopulations where we can estimate the average value of the potential
outcomes for $a=0$ and $1$ separately. For a given $a$, these subpopulations
are defined by the value of $\{e(X),\Psi_{a}(X)\}$, leading to (\ref{eq:newds-balancing}).
This difference allows us to reduce the dimension of the double score
from three to two, a small reduction of the dimension of the matching
variable, a big reduction of the bias order of the matching estimator. 

We focus on estimating $\mu_{a}=\E\{Y(a)\}$ separately for $a=0,1$.
Let the matching variable be the double score $S_{a}(X)=\{e(X),\Psi_{a}(X)\}$
or $S_{a}$ for shorthand. Table \ref{tab:Matching-schemes} (II)
illustrates the new matching scheme to impute the missing potential
outcomes. For unit $i$ with $A_{i}=1$, $\hat{Y}_{i}(1)=Y_{i}$ and
$\hat{Y}_{i}(0)=M^{-1}\sum_{l\in\J_{S_{0},i}}Y_{l}$. For unit $i'$
with $A_{i'}=0$, $\hat{Y}_{i'}(1)=M^{-1}\sum_{l\in\J_{S_{1},i'}}Y_{l}$
and $\hat{Y}_{i'}(0)=Y_{i'}$. Importantly, the new matching scheme
uses different matching variables, namely $S_{0}$ and $S_{1}$, to
impute the missing values of $Y(0)$ and $Y(1)$. This is in contrast
to matching scheme (I) that uses the same matching variable for imputing
the missing values of $Y(0)$ and $Y(1)$. Once we approximate both
potential outcomes for all units, a simple DSM estimator of $\tau$
is
\begin{equation}
\hat{\tau}_{\dsm}^{(0)}=\hat{\mu}_{1,\dsm}^{(0)}-\hat{\mu}_{0,\dsm}^{(0)},\label{eq:dsm0}
\end{equation}
 where 
\[
\hat{\mu}_{a,\dsm}^{(0)}=n^{-1}\sum_{i=1}^{n}\hat{Y}_{i}(a)=n^{-1}\sum_{i=1}^{n}\bone(A_{i}=a)\left(1+M^{-1}K_{S_{a},i}\right)Y_{i},
\]
for $a=0,1$. Following \citet{abadie2006large}, we can derive the
following decomposition
\[
n^{1/2}(\hat{\tau}_{\dsm}^{(0)}-\tau)=B_{n}+C_{n},
\]
where
\begin{eqnarray}
B_{n} & = & n^{-1/2}\sum_{i=1}^{n}(1-2A_{i})\left[M^{-1}\sum_{j\in\J_{S_{A_{i}},i}}\left\{ \mu_{A_{i}}(S_{A_{i},i})-\mu_{A_{i}}(S_{A_{i},j})\right\} \right],\label{eq:B2-dsm}\\
C_{n} & = & n^{-1/2}\sum_{i=1}^{n}\left[\mu_{1}(S_{1,i})-\mu_{0}(S_{0,i})-\tau+(2A_{i}-1)\left(1+M^{-1}K_{S_{A_{i}},i}\right)\left\{ Y_{i}-\mu_{A_{i}}(S_{A_{i},i})\right\} \right],\nonumber 
\end{eqnarray}
and $S_{a,i}=\{e(X_{i}),\Psi_{a}(X_{i})\}$, for $a=0,1$. By Lemma
\ref{lem3:newds}, we have $\E(C_{n})=0$ for $a=0,1$. Again, $B_{n}$
contributes to the asymptotic bias of the matching estimator. By Theorem
1 of \citet{abadie2006large}, $\E(B_{n})=O(n^{-1/2})$. Therefore,
$\hat{\tau}_{\dsm}^{(0)}$ is asymptotically unbiased. 

\subsection{Multiply robust DSM\label{subsec:New-multiply-robust}}

In practice, $S_{0}$ and $S_{1}$ are unknown, requiring modeling
and estimation from the observed data. Following the empirical literature,
one can posit a logistic regression model for the propensity score
and a generalized linear model for the prognostic score. To provide
additional protection against model misspecification, we can posit
multiple candidate models for both scores. The intuition is that if
at least one of the candidate models is correctly specified, whether
it is a propensity score model or a prognostic score model, balancing
at lease one score suffices to remove confounding biases. Therefore,
the DSM estimator achieves the so-called multiple robustness. 

Following \citet{han2013estimation}, we postulate multiple candidate
models 
\begin{itemize}
\item $\mathcal{M}(\alpha)=\{e^{j}(X;\alpha^{j}):j=1,\ldots,J\}$ for $e(X)$
with unknown parameters $\alpha=(\alpha^{1,\T},\ldots,\alpha^{J,\T})^{\T}$; 
\item $\mathcal{M}_{0}(\beta_{0})=\{\Psi_{0}^{k}(X;\beta_{0}^{k}):k=1,\ldots,K\}$
and $\mathcal{M}_{1}(\beta_{1})=\{\Psi_{1}^{k}(X;\beta_{1}^{k}):k=1,\ldots,K\}$
for $\Psi_{0}(X)$ and $\Psi_{1}(X),$ respectively, with unknown
parameters $\beta_{0}=(\beta_{0}^{1,\T},\ldots,\beta_{0}^{K,\T})^{\T}$
and $\beta_{1}=(\beta_{1}^{1,\T},\ldots,\beta_{1}^{K,\T})^{\T}$ . 
\end{itemize}
Let $\hat{\alpha}^{j}$, $\hat{\beta}_{0}^{k}$ and $\hat{\beta}_{1}^{k}$
be the maximum likelihood estimators or the method of moments estimators
of $\alpha^{j}$, $\beta_{0}^{k}$ and $\beta_{1}^{k}$ under the
corresponding working model, respectively. 

For each treatment level $a\in\{0,1\}$, let $S_{a}(\theta_{a})=\{\mathcal{M}(\alpha),\mathcal{M}_{a}(\beta_{a})\}$,
where $\theta_{a}^{\T}=(\alpha^{\T},\beta_{a}^{\T})$, be the the
set of candidate models for the propensity score and the prognostic
score for treatment $a$, for $a=0,1$. Under matching scheme (II),
we use $S_{a}(\hat{\theta}_{a})$ to impute the missing values of
$Y(a)$, separately for $a=0,1$. The corresponding dimension of the
matching variable is $J+K$. Let $S(\theta)=\{\mathcal{M}(\alpha),\mathcal{M}_{0}(\beta_{0}),\mathcal{M}_{1}(\beta_{1})\}$,
where $\theta=(\alpha^{\T},\beta_{0}^{\T},\beta_{1}^{\T})^{\T}$,
be the set of candidate models for the propensity score and the prognostic
score for both treatment groups. Under matching scheme (I), one would
use $S(\hat{\theta})$ as the matching variable; the corresponding
dimension is thus $J+2K$. If the number of candidate models for the
prognostic score is large, the dimension reduction of the double score
under new matching scheme (II) can be much larger than under matching
scheme (I).

The initial DSM estimator of $\tau$ is given by $\hat{\tau}_{\dsm}^{(0)}$
in (\ref{eq:dsm0}) with $S_{a}$ replaced by $S_{a}(\hat{\theta}_{a})$
for $a=0,1$. We denote the initial estimator as $\hat{\tau}_{\dsm}^{(0)}(\hat{\theta})$
to reflect its dependence on $\hat{\theta}$. As discussed in Section
\ref{subsec:New-simple-double}, if $J=K=1$, the dimension of $S_{a}(\hat{\theta})$
is two for $a=0,1$. In this case, the asymptotic bias of the matching
estimator due to matching discrepancy is negligible. We do not require
further steps to correct the asymptotic bias of $\hat{\tau}_{\dsm}^{(0)}$.
This preserves the simplicity of matching in practice. However, if
$J,K\geq2$, the dimension of each matching variable is larger than
or equal to four. Consequently, as shown in Table \ref{tab:bias order},
the bias of the matching estimator due to matching discrepancy is
not asymptotic negligible\textcolor{black}{.} In this case, we propose
the de-biasing matching estimator that corrects the asymptotic bias
due to matching discrepancy.

Let $\hat{\mu}_{a}(S_{a})$ be a nonparametric estimator of $\mu_{a}(S_{a})$,
for $a=0,1$, e.g., using the method of sieves \citep{chen2007large}.
The de-biasing DSM estimator of $\tau$ is
\begin{equation}
\hat{\tau}_{\dsm}(\hat{\theta})=\hat{\tau}_{\dsm}^{(0)}(\hat{\theta})-n^{-1/2}\hat{B}_{n},\label{eq:dsm-tau}
\end{equation}
where $\hat{B}_{n}$ is an estimator of $B_{n}$ by replacing $\mu_{a}(S_{a})$
with $\hat{\mu}_{a}(S_{a})$ for $a=0,1$.

Before delving into the discussion of the theoretical properties of
$\hat{\tau}_{\dsm}(\hat{\theta})$, we summarize the DSM algorithm
that contains nuts and bolts as follows.
\begin{description}
\item [{Step$\ 1.$}] Posit multiple candidate parametric models $\mathcal{M}(\alpha)$,
$\mathcal{M}_{0}(\beta_{0})$, and $\mathcal{M}_{1}(\beta_{1})$ for
$e(X)$, $\Psi_{0}(X)$, and $\Psi_{1}(X)$, respectively. Obtain
the parameter estimators $\hat{\alpha},$ $\hat{\beta}_{0}$, and
$\hat{\beta}_{1}$. For each unit $i$, calculate $S_{a,i}(\hat{\theta}_{a})=\{\mathcal{M}(\hat{\alpha}),\mathcal{M}_{a}(\hat{\beta}_{a})\}$
for $a=0,1$. The propensity scores are probability estimates, ranging
from zero to one. To stabilize the numerical performance, it is desirable
to use a monotone mapping to transform each propensity score estimate
$e^{j}(X_{i};\hat{\alpha}_{j})\in(0,1)$ in $\mathcal{M}(\hat{\alpha})$,
e.g., to $\logit\{e^{j}(X_{i};\hat{\alpha}_{j})\}\in\mathbb{\R}$.
We also recommend standardize $S_{a,i}(\hat{\theta}_{a})$ such that
each component has mean zero and variance one for $a=0,1$.
\item [{Step$\ 2.$}] For each unit $i$ with treatment $A_{i}=a$, find
$M$ nearest neighbors from the treatment group $1-a$ based on the
matching variable $S_{1-a,i}=S_{1-a,i}(\hat{\theta})$. \textcolor{black}{Obtain
$\J_{S_{1-a}(\hat{\theta}),i}$ that contains the indexes of the matched
units for unit $i$ and calculate $K_{S_{a}(\hat{\theta}),i}$ that
counts the number of time that unit $i$ is matched to other units.
Obtain the initial matching estimator }$\hat{\tau}_{\dsm}^{(0)}(\hat{\theta})$
in (\ref{eq:dsm0}) with $S_{a}$ replaced by $S_{a}(\hat{\theta}_{a})$.
\end{description}
If $J=K=1$, let the DSM estimator be $\hat{\tau}_{\dsm}(\hat{\theta})=\hat{\tau}_{\dsm}^{(0)}(\hat{\theta})$.
If $J,K\geq2$, we proceed to Steps 3 and 4 below. Even with $J=K=1$,
Steps 3 and 4 can help to reduce the matching discrepancy in finite
samples. 
\begin{description}
\item [{Step$\ 3.$}] Obtain a nonparametric estimator of $\mu_{a}(S_{a})$,
denoted by $\hat{\mu}_{a}(S_{a})$, e.g. by the method of sieves based
on $[\{Y_{i},S_{a,i}(\hat{\theta}_{a})\}:A_{i}=a]$, for $a=0,1$.
\item [{Step$\ 4.$}] The DSM estimator of $\tau$ is given by $\hat{\tau}_{\dsm}(\hat{\theta})$
in (\ref{eq:dsm-tau}) with $S_{a,i}$ replaced by $S_{a,i}(\hat{\theta}_{a})$. 
\end{description}

\section{Main results\label{sec:Main-results}}

In this section, we establish the asymptotic properties of $\hat{\tau}_{\dsm}(\hat{\theta})$,
which depends on the estimators of all nuisance parameters in the
propensity score and prognostic score models. Without loss of generality,
we consider the prognostic score $\Psi_{a}(X)=\mu_{a}(X;\beta_{a})$
in Example \ref{eg:location-shifted} and multiple candidate models
$\Psi_{a}^{k}(X;\beta_{a}^{k})=\mu_{a}^{k}(X;\beta_{a}^{k}),$ for
$k=1,\ldots,K$ and $a=0,1$ . Consider $\hat{\alpha}^{j}$, $\hat{\beta}_{0}^{k}$
and $\hat{\beta}_{1}^{k}$ that solve the estimating equation
\begin{equation}
n^{-1/2}\sum_{i=1}^{n}\left(\begin{array}{c}
U_{1}^{j}(A_{i},X_{i};\alpha^{j})\\
U_{2}^{k}(A_{i},X_{i},Y_{i};\beta_{0}^{k})\\
U_{3}^{k}(A_{i},X_{i},Y_{i};\beta_{1}^{k})
\end{array}\right)=0,\label{eq:pesdo score-1}
\end{equation}
where
\begin{eqnarray*}
U_{1}^{j}(A,X;\alpha^{j}) & = & \frac{\partial e^{j}(X;\alpha^{j})}{\partial\alpha^{j}}\frac{A-e^{j}(X;\alpha^{j})}{e^{j}(X;\alpha^{j})\{1-e^{j}(X;\alpha^{j})\}},\\
U_{2}^{k}(A,X,Y;\beta_{0}^{k}) & = & (1-A)\frac{\partial\mu_{0}^{k}(X;\beta_{0}^{k})}{\partial\beta_{0}^{q}}\{Y-\mu_{0}^{k}(X;\beta_{0}^{k})\},\\
U_{3}^{k}(A,X,Y;\beta_{1}^{k}) & = & A\frac{\partial\mu_{1}^{k}(X;\beta_{1}^{k})}{\partial\beta_{1}^{k}}\{Y-\mu_{1}^{k}(X;\beta_{1}^{k})\},
\end{eqnarray*}
for $j=1,\ldots,J$ and $k=1,\ldots,K$. Then, $\hat{\theta}$ solves
the joint estimating equation 
\[
\cU_{n}(\theta)=n^{-1/2}\sum_{i=1}^{n}U(A_{i},X_{i},Y_{i};\theta)=0,
\]
where $U(\theta)$ stacks $U_{1}^{j}(A_{i},X_{i},\alpha^{j})$ for
$j=1,\ldots,J$, $U_{2}^{k}(A_{i},X_{i},Y_{i};\beta_{0}^{k})$ and
$U_{3}^{k}(A_{i},X_{i},Y_{i};\beta_{1}^{k})$ for $k=1,\dots,K$. 

Let $\theta^{*}$ be the probability limit of $\hat{\theta}.$ We
divide our theoretical investigation into two steps: first, we establish
the asymptotic results for the DSM estimator when $\theta^{*}$ is
known, and second, building on the step-one results, we quantify the
impact of the estimation of $\theta^{*}$ on the asymptotic distribution. 

\subsection{Asymptotic results with known $\theta^{*}$\label{subsec:Asymptotic-results-for}}

We allow possible model misspecification, so $\theta^{*}$ may not
be the true parameter values. If $e^{j}(X;\alpha^{j})$ is a correctly
specified model, we have $e^{j}(X;\alpha^{j*})=e(X)$; if $\mu_{a}^{k}(X;\beta_{a}^{k})$
is a correctly specified model, we have $\mu_{a}^{k}(X;\beta_{a}^{k*})=\mu_{a}(X)$,
for $a=0,1$. The key insight is that if any model of the propensity
score or prognostic score is correctly specified, $S_{a}(\theta^{*})$
remains as a balancing score in the sense that $Y(a)\indep A\mid S_{a}(\theta^{*})$
holds for $a=0,1$ (Lemma \ref{lem3:newds}). Based on this key observation,
we will show that the DSM estimator is multiply robust. 

Before presenting the asymptotic properties of $\hat{\tau}_{\dsm}(\theta^{*})$,
we require technical conditions. For simplicity, let $S_{a}=S_{a}(\theta^{*})$
and let $f_{1}(S_{a})$ and $f_{0}(S_{a})$ be the conditional density
of $S_{a}$ given $A=1$ and $A=0$, respectively.

\begin{assumption}\label{asmp:tau}For $a=0,1$, (i) the matching
variable $S_{a}$ has a compact and convex support $\mathcal{S}$,
with a continuous density bounded and bounded away from zero: there
exist constants $C_{1L}$ and $C_{1U}$ such that $C_{1L}\leq f_{1}(S_{a})/f_{0}(S_{a})\leq C_{1U}$
for any $S_{a}\in\mathcal{S}$; (ii) $\mu_{a}(S_{a})$ and $\sigma_{a}^{2}(S_{a})$
satisfy Lipschitz continuity conditions: there exists a constant $C_{2}$
such that $|\mu_{a}(S_{a,i})-\mu_{a}(S_{a,j})|<C_{2}||S_{a,i}-S_{a,j}||$
for any $S_{a,i}$ and $S_{a,j}$, and similarly for $\sigma_{a}^{2}(S_{a})$;
and (iii) there exists $\delta>0$ such that $\E\left\{ |Y(a)|^{2+\delta}\mid S_{a}\right\} $
is uniformly bounded for any $S_{a}\in\mathcal{S}.$

\end{assumption}

Assumption \ref{asmp:tau} has been considered by \citet{abadie2006large}
and \citet{abadie2016matching} for matching estimators based on the
covariates and the propensity score. Assumption \ref{asmp:tau} (i)
a convenient regularity condition. \textcolor{black}{Assumption \ref{asmp:tau}
(ii) imposes smoothness conditions for the outcome mean function $\mu_{a}(S_{a})$
and the variance function $\sigma_{a}^{2}(S_{a})$.} Assumption \ref{asmp:tau}
(iii) is a moment condition for establishing the central limit theorem.

In the following theorem, we establish the multiple robustness and
asymptotic distribution of $\hat{\tau}_{\dsm}(\theta^{*})$.

\begin{theorem}\label{Thm:1}Under Assumptions \ref{asump-overlap}\textendash \ref{asmp:tau},
if any model of the propensity score or prognostic score is correctly
specified, we have $n^{1/2}\left\{ \hat{\tau}_{\dsm}(\theta^{*})-\tau\right\} \rightarrow\N(0,V_{\tau}),$
in distribution, as $n\rightarrow\infty$, where
\begin{multline}
V_{\tau}=\E\left[\{\mu_{1}(S_{1})-\mu_{0}(S_{0})-\tau\}^{2}\right]+\E\left(\sigma_{1}^{2}(S_{1})\left[\frac{1}{e(S_{1})}+\frac{1}{2M}\left\{ \frac{1}{e(S_{1})}-e(S_{1})\right\} \right]\right)\\
+\E\left(\sigma_{0}^{2}(S_{0})\left[\frac{1}{1-e(S_{0})}+\frac{1}{2M}\left\{ \frac{1}{1-e(S_{0})}-1+e(S_{0})\right\} \right]\right).\label{eq:V1}
\end{multline}

\end{theorem}

It is worth comparing the three matching methods namely PSM, PGM
and DSM based on the variance formula in (\ref{eq:V1}). To simplify
the discussion, we consider one propensity score model and one prognostic
score model. We show in the supplementary material that \textit{if
the prognostic score model is correctly specified}, the DSM estimator
may be less efficient than the PGM estimator of $\tau$. \textit{If
the propensity score model is correctly specified}, the DSM estimator
is more efficient than the PSM estimator of $\mu_{a}$ for $a=0,1$;
however, this improvement is not guaranteed for estimating $\tau$.
Importantly, DSM has the advantage of \textit{double robustness} compared
to single score matching: the DSM estimator of $\tau$ is consistent
if either model of the propensity score or prognostic score is correctly
specified, but not necessarily both. Moreover, the consistency is
agnostic to which model is correctly specified. The multiple model
specification offers additional protection against model misspecification.

From Theorem \ref{Thm:1}, the consistency of the DSM estimator is
guaranteed if any model for the propensity score or prognostic score
is correctly specified. Both the number of the posited models and
their functional forms can affect the efficiency of the DSM estimator
in a very complex way. In addition, with a finite sample size, the
matching performance can be unstable if there are a large number of
working models. In particular, the discrepancy of the matched units
may be large when some of the models are poorly constructed. To reduce
the chance of running into these issues, we suggest positing a few
well-constructed working models instead of a large number of poorly
built ones.

\subsection{Asymptotic results with estimated $\theta^{*}$\label{subsec:Asymptotic-results-for-1}}

To acknowledge the fact that $\theta^{*}$ is estimated prior to matching,\textcolor{black}{{}
}we will establish the approximate distribution of $\hat{\tau}_{\dsm}(\hat{\theta})$
and examine the impact of nuisance parameter estimation on the properties
of the DSM estimator.\textcolor{black}{{} As in \citet{abadie2016matching},}
the typical Taylor expansion technique can not be used because of
the non-smooth nature of matching. Our derivation is based on the
technique developed by \citet{andreou2012alternative}, which offers
a general approach for deriving the limiting distribution of statistics
that involve estimated nuisance parameters. \textcolor{black}{This
technique has been successfully used by \citet{abadie2016matching}
for the PSM estimators of the ATE and ATT based on} a correctly specified
propensity score model. We extend their results to the DSM estimator
requiring only one of the double score models to be correctly specified.

The main theorem of \textcolor{black}{\citet{abadie2016matching}}
is the application of Le Cam's third lemma (see Section \ref{sec:Le-Cam's-third})
to Locally Asymptotically Normal (LAN) models. Let $\bP^{\theta^{*}}$
be the true probability measure of $n$ copies of the random variables,
$\theta_{n}$ contiguous to $\theta^{*}$, and $\bP^{\theta_{n}}$
the probability measure with the local parameter $\theta_{n}$. Assuming
that under $\bP^{\theta_{n}}$,
\[
\{n^{1/2}\{\hat{\tau}_{\dsm}(\theta_{n})-\tau(\theta_{n})\},n^{1/2}(\hat{\theta}-\theta_{n}),\log(\de\mathbb{P}^{\theta^{*}}/\de\mathbb{P}^{\theta_{n}})\}^{\T}
\]
has a limiting Normal distribution. By Le Cam\textquoteright s third
lemma, under $\mathbb{P}^{\theta^{*}}$, $n^{1/2}\{\hat{\tau}_{\dsm}(\theta_{n})-\tau(\theta^{*})\}$
has a limiting Normal distribution. Heuristically, by replacing $\theta_{n}$
with $\hat{\theta}$, one can then approximate the asymptotic distribution
of $n^{1/2}\{\hat{\tau}_{\dsm}(\hat{\theta})-\tau\}$. 

Under a correctly specified propensity score model,\textcolor{black}{{}
$\mathbb{P}^{\theta^{*}}$ is naturally the probability measure governed
by the likelihood function of $\theta^{*}$. In our setting, we posit
multiple working models for the propensity score and prognostic score
and require only one model to be correctly specified. In this case,
it is difficult to characterize $\mathbb{P}^{\theta^{*}}$. }Our key
step is to recognize that $\theta^{*}$ is defined based on $\E\{U(\theta^{*})\}=0$,
which entails a semiparametric model with mean restrictions. To invoke
the Le Cam's lemma,\textcolor{black}{{} we} consider a semiparametric
model for $\theta^{*}$ based on the asymptotic distribution of the
estimating function $\cU_{n}(\theta^{*})$. We then carry over the
inferential framework of to our context. 

\begin{theorem} \label{Thm:3} Under Assumptions \ref{asump-overlap}\textendash \ref{asmp:tau},
and regularity conditions specified in the supplementary material,
if any model of the propensity score or prognostic score is correctly
specified, the approximate distribution of $n^{1/2}\left\{ \hat{\tau}_{\dsm}(\hat{\theta})-\tau\right\} $
is $\N(0,V_{\tau,\adj}),$ where
\begin{equation}
V_{\tau,\adj}=V_{\tau}-\gamma_{1}^{\T}\Sigma_{U}^{-1}\gamma_{1}+\gamma_{2}^{\T}\Sigma_{\theta^{*}}\gamma_{2},\label{eq:sig2_adj}
\end{equation}
where $V_{\tau}$ is given in (\ref{eq:V1}), $\Sigma_{U}=\E\{U(A,X,Y;\theta^{*})$
$U(A,X,Y;\theta^{*})^{\T}\}$, $\Sigma_{\theta^{*}}=\Gamma_{\theta^{*}}^{-1}\Sigma_{U}(\Gamma_{\theta^{*}}^{-1})^{\T}$,
$\Gamma_{\theta^{*}}=\E\{\partial U(A,X,$ $Y;\theta^{*})/\partial\theta^{\T}\}$,
$\gamma_{1}$ and $\gamma_{2}$ are given in (\ref{eq:gamma1}) and
(\ref{eq:gamma2}), respectively. 

\end{theorem}

We discuss the impact of estimating the nuisance parameters on the
matching estimators.\textcolor{black}{{} \citet{abadie2016matching}
showed that for $\tau$, matching on the estimated propensity score
always improves the estimation efficiency compared to matching on
the true propensity score.} This improvement is due to the correlation
of the matching estimator and the score function for the parameters
in the propensity score. In our context, comparing the asymptotic
variances in Theorems \ref{Thm:1} and \ref{Thm:3}, the difference
between $V_{\tau,\adj}$ and $V_{\tau}$, $-\gamma_{1}^{\T}\Sigma_{U}^{-1}\gamma_{1}+\gamma_{2}^{\T}\Sigma_{\theta^{*}}\gamma_{2}$,
can be either positive, negative, or zero; i.e.,\textcolor{black}{{}
matching on the estimated double score can either increase, decrease,
or maintain the estimation efficiency compared to matching on the
true double score. To explain the difference,} we note that the variance
reduction term $-\gamma_{1}^{\T}\Sigma_{U}^{-1}\gamma_{1}$ is still
due to the correlation of the matching estimator and the score function
for the parameters in the double score, while the variance inflation
term $\gamma_{2}^{\T}\Sigma_{\theta^{*}}\gamma_{2}$ is because if
either the prognostic score model or the propensity score model is
misspecified, $\tau$ may depend on the nuisance parameters through
$\tau=\E\left[\mu_{1}\{S_{1}(\theta^{*})\}-\mu_{0}\{S_{0}(\theta^{*})\}\right],$
which contributes to the variance inflation term. On the other hand,
\textcolor{black}{\citet{abadie2016matching} focused on the setting
when the propensity score model is the only nuisance model and is
correctly specified. In this case, $\tau$ does not depend on $\alpha^{*}$,
$\gamma_{2}$ is zero, and therefore the variance inflation term is
zero. }

\subsection{Variance estimation and inference\label{sec:Variance-estimation}}

Theorem \ref{Thm:3} provides a guidance for variance estimation of
the DSM estimators that can take all sources of variability into account.
However, such variance estimators are complicated to construct. We
consider variance estimation based on replication methods \citep{efron1979bootstrap,wolter2007introduction}.
Lack of smoothness makes the standard replication methods invalid
for the matching estimator. When the number of matches remains fixed,
\textcolor{black}{\citet{abadie2008failure} demonstrated the failure
of the bootstrap for matching estimators. This is because the non-parametric
bootstrap cannot preserve the distribution of the number of times
that each unit is used as a match. In this case,} \citet{otsu2016bootstrap}
proposed a wild bootstrap procedure for the matching estimator when
matching is directly based on the covariates. \citet{yang2017predictive}
proposed a replication based procedure for predictive mean matching
in survey data. 

Given the two-stage estimation procedure for the DSM estimator, the
variability of the matching estimator results from two sources: first,
the estimation of the double score function, and second, matching.
To faithfully take into account all sources of variability, we propose
a two-stage replication variance estimation procedure, in parallel
to the two-stage point estimation procedure. First, we construct replicates
of the nuisance parameter estimators in the double score. Second,
based on the asymptotic linear representations of the DSM estimator,
we construct replicates of the DSM estimator directly based on the
linear terms with the replicated nuisance parameters. In this way,
the distribution of the number of times that each unit is used as
a match can be retained.

Specifically, the replication variance estimation algorithm proceeds
as follows.
\begin{description}
\item [{VE-Step$\ 1.$}] Obtain a bootstrap sample, or equivalently the
bootstrap replication weights $\omega_{i}^{*}=n^{-1}m_{i}^{*}$ with
$(m_{1}^{*},\ldots,m_{n}^{*})$ is a multinomial random vector with
$n$ draws on $n$ equal probability cells. Obtain a bootstrap replicate
of $\hat{\theta}$, $\hat{\theta}^{*}$, by solving the estimating
equation $n^{-1/2}\sum_{i=1}^{n}\{\omega_{i}^{*}U(A_{i},X_{i},Y_{i};\theta)\}=0$.
For each unit $i$, calculate $S_{a,i}(\hat{\theta}^{*})$ for $a=0,1$.
\item [{VE-Step$\ 2.$}] Obtain a bootstrap replicate of $\hat{\tau}_{\dsm}(\hat{\theta})$
as
\begin{eqnarray*}
\hat{\tau}_{\dsm}^{*}(\hat{\theta}^{*}) & = & n^{-1}\sum_{i=1}^{n}\omega_{i}^{*}\left[\hat{\mu}_{1}\{S_{1,i}(\hat{\theta}^{*})\}-\hat{\mu}_{0}\{S_{0,i}(\hat{\theta}^{*})\}\right]\\
 &  & +n^{-1}\sum_{i=1}^{n}\omega_{i}^{*}(2A_{i}-1)\left\{ 1+M^{-1}K_{S_{A_{i}}(\hat{\theta}),i}\right\} \left[Y_{i}-\hat{\mu}_{A_{i}}\{S_{A_{i},i}(\hat{\theta}^{*})\}\right].
\end{eqnarray*}
\item [{VE-Step$\ 3.$}] Repeat VE-Steps 1 and 2 for a large number of
times. Calculate the bootstrap variance estimator of $\hat{\tau}_{\dsm}(\hat{\theta})$
as the empirical variance of $\hat{\tau}_{\dsm}^{*}(\hat{\theta}^{*})$
over a large number of bootstrap replicates.
\end{description}
\begin{remark}In VE-Step1, instead of generating bootstrap resamples,
we can generate replication weights from general distributions that
satisfy $\E(\omega_{i}^{*}\mid\obs)=1$, $\E(\omega_{i}^{*2}\mid\obs)=1$,
and $\E(\omega_{i}^{*4}\mid\obs)<\infty$, where 'obs' denotes the
observed data. For example, one can generate the $\omega_{i}^{*}$'s
from Exp$(1)$ independently from the observed data.

\end{remark}

\section{Multiply robust matching estimator of the QTE\label{sec:QTE}}

Matching is attractive for general causal estimation, because it can
be viewed as a hot deck imputation method \citep{ford1983overview},
where for each unit the donors for the missing potential outcome are
actually observed values from the opposite treatment group. An advantage
of hot deck imputation is that it preserves the distribution of the
potential outcomes so that valid estimators for parameters depending
on the entire distribution of the potential outcomes such as the mean
and quantiles can be obtained based on the imputed data set. In this
section, we extend the DSM framework to estimate the QTE. 

We focus on estimating $q_{a,\xi}$ separately for $a=0,1$. Similar
to (\ref{eq:newds-balancing}), we have

\[
q_{a,\xi}=\inf_{q}\left(\E[\pr\{Y\leq q\mid A=a,e(X),\Psi_{a}(X)\}]\geq\xi\right).
\]
Based on the above equation, we propose the DSM estimator of $q_{a,\xi}$
as
\begin{equation}
\hat{q}_{a,\xi,\dsm}=\inf_{q}\{\hat{F}_{a,\dsm}(q)\geq\xi\},\label{eq:dsm-qte}
\end{equation}
where
\begin{eqnarray}
\hat{F}_{a,\dsm}(q) & = & \hat{F}_{a,\dsm}^{(0)}(q)-n^{-1/2}\hat{B}_{a,n}(q),\label{eq:F_a,dsm}\\
\hat{F}_{a,\dsm}^{(0)}(q) & = & n^{-1}\sum_{i=1}^{n}\bone(A_{i}=a)\left(1+M^{-1}K_{S_{a},i}\right)\bone(Y_{i}\leq q),\nonumber \\
\hat{B}_{a,n}(q) & = & -n^{-1/2}\sum_{i=1}^{n}\bone(A_{i}=1-a)M^{-1}\sum_{j\in\J_{S_{a},i}}\left\{ \hat{F}_{a}(q;S_{a,i})-\hat{F}_{a}(q;S_{a,j})\right\} ,\label{eq:qte1-1}
\end{eqnarray}
and $\hat{F}_{a}(q;S_{a})$ is a semi/nonparametric estimator of $F_{a}(q;S_{a})=\pr\{Y(a)\leq q\mid S_{a}\}$,
for $a=0,1$. Note that $\hat{F}_{a,\dsm}^{(0)}(q)$ is an initial
matching estimator of $F_{a}(q)=\bP\{Y(a)\leq q\}$ and $\hat{B}_{a,n}(q)$
is the bias correction term; see the supplementary material. Then
the DSM estimator of $\Delta_{\xi}$ is $\hat{\Delta}_{\xi,\dsm}=\hat{q}_{1,\xi,\dsm}-\hat{q}_{0,\xi,\dsm}$.

For estimating $\Delta_{\xi}$, Steps $1$ and $2$ of DSM in Section
\ref{subsec:New-multiply-robust} remain the same; Steps $3'$ and
$4'$ proceed as follows:
\begin{description}
\item [{Step$\ 3'.$}] Obtain a semiparametric estimator of $F_{a}(q;S_{a})$
based on $[\{Y_{i},S_{a,i}(\hat{\theta})\}:A_{i}=a]$, for $a=0,1$.
For example, we can consider the method of sieves for the normal linear
model after a Box-Cox transformation of \citet{zhang2012causal} or
the single-index conditional distribution model of \citet{chiang2012new}.
\item [{Step$\ 4'.$}] The DSM estimator of $q_{a,\xi}$ is given by (\ref{eq:dsm-qte})
with $S_{a}$ replaced by $S_{a}(\hat{\theta})$. We denote the final
estimator of $q_{a,\xi}$ as $\hat{q}_{a,\xi\dsm}(\hat{\theta})$
to reflect its dependence on $\hat{\theta}$, for $a=0,1$. Then,
the DSM estimator of $\Delta_{\xi}$ is $\hat{\Delta}_{\xi,\dsm}(\hat{\theta})=\hat{q}_{1,\xi,\dsm}(\hat{\theta})-\hat{q}_{0,\xi,\dsm}(\hat{\theta})$.
\end{description}
To establish the multiple robustness and asymptotic distributions
of $\hat{\Delta}_{\xi,\dsm}(\theta^{*})$ and $\hat{\Delta}_{\xi,\dsm}(\hat{\theta})$,
we require further technical conditions.

\begin{assumption}\label{asmp:qte}For $a=0,1$, the following conditions
hold for the parameter $q_{a,\xi}$ and the estimating function $F_{a}(q)$:
(i) $q_{a,\xi}$ lies in a closed interval $\It$; (ii) the estimating
equation $F_{a}(q)=\xi$ has a unique root in the interior of $\It$;
(iii) $F_{a}(q)$ is strictly increasing and absolutely continuous
with finite first derivative in $\It$, and the derivative $f_{a}(q)=\de F_{a}(q)/\de q$
is bounded away from $0$ for all $q$ in $\It$; and (iv) for $a=0,1,$
$F_{a}(q;S_{a})$ satisfies a Lipschitz continuity condition: there
exists a constant $C_{3}$ such that $|F_{a}(q;S_{a,i})-F_{a}(q;S_{a,j})|<C_{2}||S_{a,i}-S_{a,j}||$
for any $S_{a,i}$ and $S_{a,j}$.

\end{assumption}

Under Assumptions \ref{asump-overlap}\textendash \ref{asump-ignorable},
\ref{asmp:tau} (i) and \ref{asmp:qte}, if any model of the propensity
score or prognostic score is correctly specified, similar to the proof
in Section \ref{sec:Proof-of-Theorem1}, we have $\de\hat{F}_{a,\dsm}(q_{a,\xi})/\de q=f_{a}(q_{a,\xi})+o_{\mathrm{P}}(n^{-1/2})$,
and then we express $\hat{q}_{a,\xi,\dsm}$ as
\begin{equation}
\hat{q}_{a,\xi,\dsm}-q_{a,\xi}=-\frac{\hat{F}_{a,\dsm}(q_{a,\xi})-F_{a}(q_{a,\xi})}{f_{a}(q_{a,\xi})}+o_{P}(n^{-1/2}).\label{eq:quantile nni}
\end{equation}
Expression (\ref{eq:quantile nni}) is called the Bahadur-type representation
for $\hat{q}_{a,\xi,\dsm}$ \citep{francisco1991quantile}. With the
representation (\ref{eq:quantile nni}), it is straightforward to
extend the multiple robustness and asymptotic distributions of the
ATE estimation to $\hat{\Delta}_{\xi,\dsm}(\theta^{*})$ and $\hat{\Delta}_{\xi,\dsm}(\hat{\theta})$. 

\begin{theorem}\label{Thm:2}Under Assumptions \ref{asump-overlap}\textendash \ref{asump-ignorable},
\ref{asmp:tau} (i) and \ref{asmp:qte}, if any model of the propensity
score or prognostic score is correctly specified, $n^{1/2}\left\{ \hat{\Delta}_{\xi,\dsm}(\theta^{*})-\Delta_{\xi}\right\} \rightarrow\N(0,V_{\Delta}),$
in distribution, as $n\rightarrow\infty$, where $V_{\Delta}$ is
given in (\ref{eq:V2}).

\end{theorem}

\begin{theorem} \label{Thm:3-q} Under Assumptions \ref{asump-overlap}\textendash \ref{asump-ignorable},
\ref{asmp:tau} (i) and \ref{asmp:qte}, and regularity conditions
specified in the supplementary material, if any model of the propensity
score or prognostic score is correctly specified, the approximate
distribution of $n^{1/2}\left\{ \hat{\Delta}_{\xi,\dsm}(\theta^{*})-\Delta_{\xi}\right\} $
is $\N(0,V_{\Delta,\adj}),$ where 
\begin{equation}
V=V_{\Delta}-\gamma_{3}^{\T}\Sigma_{U}^{-1}\gamma_{3}+\gamma_{4}^{\T}\Sigma_{\theta^{*}}\gamma_{4},\label{eq:sig2_adj-1}
\end{equation}
$V_{\Delta}$ is given in (\ref{eq:V2}), $\Sigma_{U}$ and $\Sigma_{\theta^{*}}$
are given in Theorem 2, $\gamma_{3}$ and $\gamma_{4}$ are given
in (\ref{eq:gamma3}) and (\ref{eq:gamma4}), respectively.

\end{theorem}

For variance estimation of $\hat{\Delta}_{\xi,\dsm}(\hat{\theta})$,
VE-Step 1 in Section \ref{sec:Variance-estimation} remains the same;
VE-Steps 2 and 3 proceed as follows:
\begin{description}
\item [{VE-Step$\ 2'.$}] For $a=0,1$, obtain a bootstrap replicate of
$\hat{q}_{a,\xi\dsm}(\hat{\theta})$, $\hat{q}_{a,\xi\dsm}^{*}(\hat{\theta}^{*})$,
by solving 
\begin{multline*}
\hat{F}_{a,\dsm}^{*}(q)=n^{-1}\sum_{i=1}^{n}\omega_{i}^{*}\hat{F}_{a}\{q;S_{a,i}(\hat{\theta}^{*})\}\\
+n^{-1}\sum_{i=1}^{n}\omega_{i}^{*}\bone(A_{i}=a)\left\{ 1+M^{-1}K_{S_{a}(\hat{\theta}),i}\right\} \left[\bone(Y_{i}\leq q)-\hat{F}_{a}\{q;S_{a,i}(\hat{\theta}^{*})\}\right]=\xi,
\end{multline*}
for $q$. Then a bootstrap replicate of $\hat{\Delta}_{\xi,\dsm}(\hat{\theta})$
is $\hat{\Delta}_{\xi,\dsm}^{*}(\hat{\theta}^{*})=\hat{q}_{1,\xi,\dsm}^{*}(\hat{\theta}^{*})-\hat{q}_{0,\xi,\dsm}^{*}(\hat{\theta}^{*})$.
\item [{VE-Step$\ 3'.$}] Repeat VE-Steps 1 and 2' for a large number of
times. Calculate the bootstrap variance estimator of $\hat{\Delta}_{\xi,\dsm}(\hat{\theta})$
as the empirical variance of $\hat{\Delta}_{\xi,\dsm}^{*}(\hat{\theta}^{*})$
over a large number of bootstrap replicates.
\end{description}

\section{Simulation\label{sec:Simulation-study}}

We conduct a simulation study to investigate the finite-sample performance
of the proposed DSM estimators relative to existing weighting and
matching estimators. In the causal inference and missing data literature,
previous simulations (e.g., \citealp{kang2007demystifying}) have
found that weighting estimators can have high variability, especially
if the probabilities are close to zero or one. \citet{frolich2004finite}
found that the weighting estimator was inferior to matching estimators
in terms of root mean squared error. It has been found that matching
on high-dimensional covariates is not practical for commonly found
sample sizes (e.g., \citealp{abadie2006large}). In the comparative
effectiveness research, PGM has been shown to be more advantageous
than PSM when the propensity score distributions are strongly separated
\citep{wyss2015matching,kumamaru2016dimension}. \citet{imbens2004nonparametric}
noted that if the regression models are misspecified, PGM may be inconsistent.
These results motivate us to compare the weighting and matching estimators
in a setting with complex data generative models, and where the propensity
scores may be close to zero or one. 

Let the sample size be $n=1000.$ Confounder $X\in\mathbb{R}^{10}$
is generated by $X_{j}\stackrel{\text{iid}}{\sim}$ Uniform {[}$1-\sqrt{3},1+\sqrt{3}${]}
for $j=1,\ldots,10$. To introduce nonlinear relationships between
$X$ and dependent variables, let $Z\in\mathbb{R}^{10}$ be a nonlinear
transformation of $X$, where $Z_{1}=\exp(X_{1}/2)$, $Z_{2}=\exp(X_{2}/3)$,
$Z_{3}=\log\{(X_{3}+1)^{2}\}$, $Z_{4}=\log\{(X_{4}+1)^{2}\}$, $Z_{5}=\bone(X_{5}>0.5)$,
$Z_{6}=\bone(X_{6}>0.75)$, $Z_{7}=\sin(X_{7}-X_{8})$, $Z_{8}=\cos(X_{7}+X_{8})$,
$Z_{9}=\sin(X_{9})$, $Z_{10}=\cos(X_{10})$, which are further scaled
and centered such that $\E(Z_{j})=1$ and $\V(Z_{j})=1$ for all $j$.
The potential outcomes are $Y(0)=\beta_{0}^{\T}Z+\epsilon(0)$ and
$Y(1)=Y(0)-\epsilon(0)+\epsilon(1)$, where $\beta_{0}^{\T}=(1,1,1,1,1,-1,-1,-1,-1,-1)/2$,
$\epsilon(0)\sim\mathcal{N}(0,2^{2})$ and $\epsilon(1)\sim\mathcal{N}(0,1)$.
Under the data generative model, the ATE $\tau$ is $0$ and the $75$th
QTE is $-0.45$. The treatment indicator $A$ follows Bernoulli$\{e(X)\}$,
where logit$\{e(X)\}=\alpha_{0}^{\T}Z$, where $\alpha_{0}^{\T}=(1,1,1,1,1,-1,-1,-1,-1,-1)/4$.
Some values of $e(X)$ are close to zero or one. 

To assess the multiple robustness property of the DSM estimators,
we consider two model specifications for the propensity score: 1)
a correctly specified logistic regression model $\logit\{e^{1}(X;\alpha^{1})\}=\alpha^{1,\T}Z$
and 2) a misspecified logistic regression model $\logit\{e^{2}(X;\alpha^{2})\}=\alpha^{2,\T}X$;
we also consider two model specifications for the prognostic score:
3) a correctly specified regression model $\mu_{a}^{1}(X;\beta_{a}^{1})=\beta_{a}^{1,\T}Z$
for $a=0,1$ and 4) a misspecified regression model $\mu_{a}^{2}(X;\beta_{a}^{2})=\beta_{a}^{2,\T}X$
for $a=0,1$.

We compare the following estimators: 
\begin{enumerate}
\item naive, which is the simple difference of standard estimators from
two treatment groups; 
\item the weighting estimators including the IPW and AIPW estimators (``ipw''
and ``aipw'');
\item the matching estimators based on $X$ (``m.x''; bias-corrected \citealp{abadie2011bias}),
or propensity score (``psm'') or prognostic score (``pgm'') or
double score (``dsm''). 
\end{enumerate}
Each weighting and matching estimator is assigned a name in the form
of ``method-0000'', where each digit of the four-digit number, from
left to right, indicates if $e^{1}(X;\alpha^{1})$, $e^{2}(X;\alpha^{2})$,
$\{\mu_{a}^{1}(X;\beta_{a}^{1})\}_{a=0}^{1}$, or $\{\mu_{a}^{2}(X;\beta_{a}^{2})\}_{a=0}^{1}$
is used in the construction of the method, with ``1'' meaning yes
and ``0'' meaning no, respectively. For example, ``ipw1000'' is
the IPW estimator with the propensity score model $e^{1}(X;\alpha^{1})$
and ``dsm1110'' is the DSM estimator with two propensity score models
$e^{1}(X;\alpha^{1})$, $e^{2}(X;\alpha^{2})$ and one prognostic
score model $\{\mu_{a}^{1}(X;\beta_{a}^{1})\}_{a=0}^{1}$. We implement
standard IPW and AIPW estimators for the ATE estimation and the corresponding
estimators of \citet{zhang2012causal} for the QTE estimation. For
all matching estimators, the conditional outcome mean functions are
approximated using power series; and the conditional distribution
functions are approximated based on the power series for the Normal
linear model \citet{zhang2012causal}.

\begin{figure}
\begin{centering}
\includegraphics[scale=0.65]{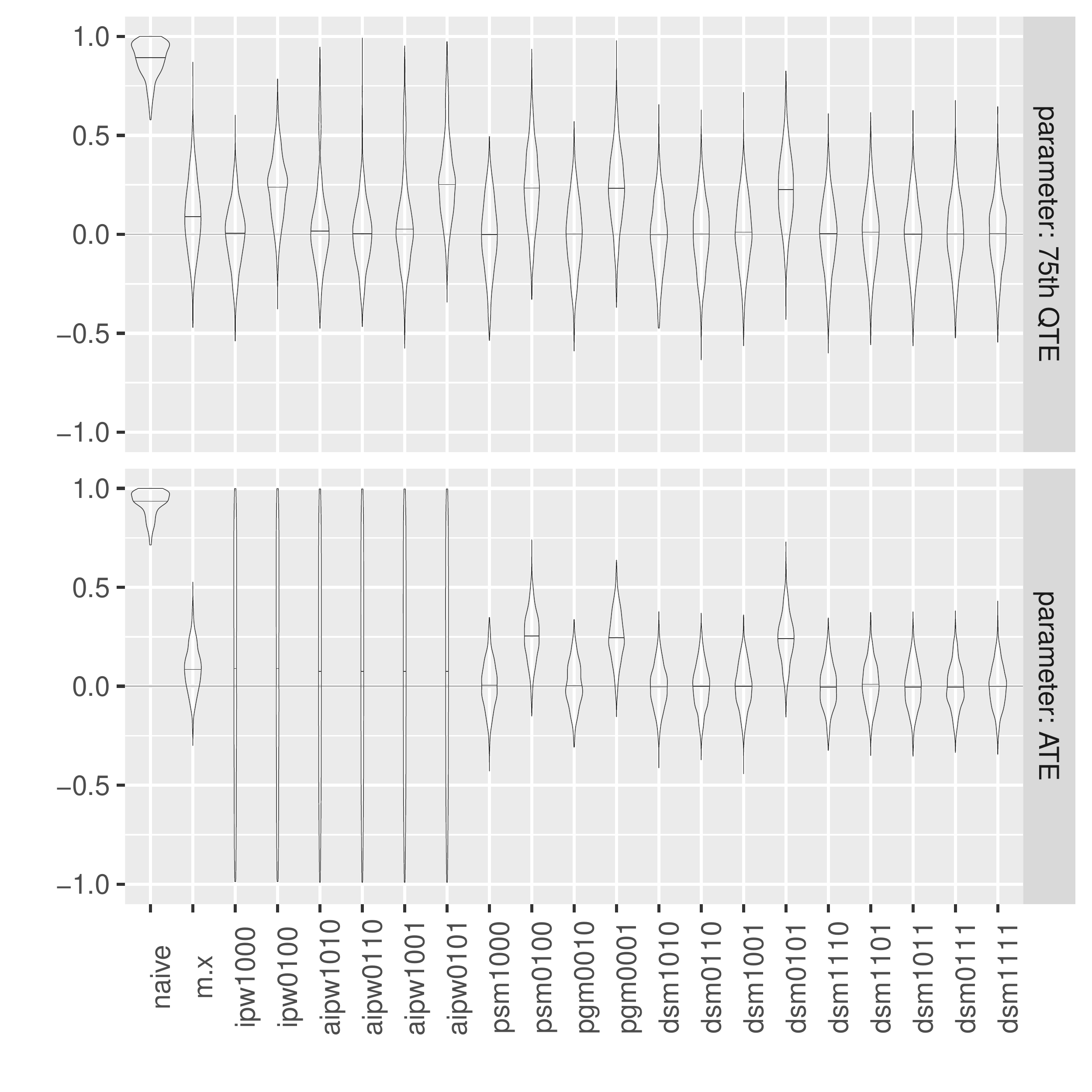}
\par\end{centering}
\caption{\label{fig1:sim}Simulation results of various weighting and matching
estimators. There are two panels of results: the top for the $75$th
QTE and the bottom for the ATE. Each violin plot shows the distribution
of the estimator subtracting the true parameter value based on $1000$
Monte Carlo simulated datasets. Due to the extreme estimates of the
weighting estimators of the ATE, the violin plots for ``ipwXXXX''
and ``aipwXXXX'' (with each X being $0$ or $1$) are truncated
at $-1$ and $1$ and appear to be slim.}
\end{figure}
Figure~\ref{fig1:sim} shows the distributions of the estimation
error (i.e.; the estimator minus the true parameter value) based on
1000 repeated sampling. The naive estimator is biased for the $75$th
QTE and the ATE. Matching directly based on $10$-dimensional $X$
(indicated by ``m.x'') is biased for the QTE and the ATE even with
bias correction. This suggests that matching on high-dimensional covariates
is not practical and calls for dimension reduction. We discuss the
results from the IPW and AIPW estimator separately for the QTE and
ATE estimation. For the QTE estimation, the IPW estimator relies on
a correct specification of the propensity score: it has small biases
if the propensity score model is correctly specified (indicated by
``ipw1000''); while it is biased if the propensity score model is
misspecified (indicated by ``ipw0100''). The AIPW estimator is doubly
robust: it has small biases if either the propensity score model or
the prognostic model is correctly specified (indicated by ``aipw1010'',
``aipw1001'', ``aipw0110''). These performances are expected based
on the existing results for the IPW and AIPW estimators. Surprisingly,
the IPW and AIPW estimators are biased and have very large variability
for estimating the ATE, even when the involving models are correctly
specified. We examine the empirical distribution of the estimated
propensity score weights and find that there are extremely large weights
that dominate other weights. To mitigate this issue, one can stabilize
the weighting estimators by normalizing the weights \citep{hernan2001marginal}.
However, we do not find this strategy effective in our setting. Although
the AIPW estimator is constructed to be semiparametrically efficient,
its performance can be severely poor when it involves large weights.
By construction, matching does not invert the estimated propensity
scores and therefore is more robust to outliers of the propensity
score estimates. We now compare performances of the score-based matching
estimators. The single score matching estimators (indicated by ``psm1000'',
``psm0100'', ``pgm0010'', ``pgm0001'') are singly robust and
rely on a correct specification of the underlying score model. The
DSM estimators are multiply robust in that it has small biases for
the QTE and the ATE if any model of the propensity score or prognostic
score is correctly specified. Unlike the AIPW estimators, the DSM
estimators are robust to extreme values of the propensity score estimates.
Table \ref{tab:Sim1} reports the coverage rates of the DSM estimators
of the $75$th QTE and the ATE using the proposed replication-based
method. Under the multiple robustness condition (i.e., if any model
of the propensity score or prognostic score is correctly specified),
the coverage rates are all close to the nominal coverage except for
``dsm0101''. 

\begin{table}
\begin{centering}
{\scriptsize{}\caption{{\scriptsize{}\label{tab:Sim1}}Simulation results based on $1000$
Monte Carlo simulated datasets for the coverage properties of the
DSM estimators using the replication-based method: empirical coverage
rate and (empirical coverage rate $\pm$ $1.96\times$Monte Carlo
standard error)}
}{\scriptsize\par}
\par\end{centering}
\centering{}
\centering{}%
\begin{tabular}{lcc||ccc||c}
\hline 
 & \multicolumn{3}{c}{75th QTE} & \multicolumn{3}{c}{ATE}\tabularnewline
\cline{2-7} \cline{3-7} \cline{4-7} \cline{5-7} \cline{6-7} \cline{7-7} 
``dsm1010'' & 95.4 & \multicolumn{2}{c}{(94.6, 97.2)} & 95.6 & \multicolumn{2}{c}{(94.4, 96.9)}\tabularnewline
``dsm0110'' & 96.5 & \multicolumn{2}{c}{(95.2, 97.1)} & 95.9 & \multicolumn{2}{c}{(94.7, 97.1)}\tabularnewline
``dsm1001'' & 96.6 & \multicolumn{2}{c}{(95.3, 98.2)} & 96.0 & \multicolumn{2}{c}{(94.9,97.1)}\tabularnewline
``dsm0101'' & 79.7 & \multicolumn{2}{c}{(77.2, 82.2)} & 55.1 & \multicolumn{2}{c}{(52.0, 58.2)}\tabularnewline
``dsm1111'' & 95.4 & \multicolumn{2}{c}{(94.1, 96.7)} & 95.6 & \multicolumn{2}{c}{(94.3, 96.9)}\tabularnewline
``dsm1110'' & 95.9 & \multicolumn{2}{c}{(94.7, 97.1)} & 96.1 & \multicolumn{2}{c}{(94.9, 97.3)}\tabularnewline
``dsm1101'' & 95.7 & \multicolumn{2}{c}{(94.5, 96.9)} & 95.5 & \multicolumn{2}{c}{(94.3, 96.9)}\tabularnewline
``dsm1011'' & 95.0 & \multicolumn{2}{c}{(93.6, 96.4)} & 95.5 & \multicolumn{2}{c}{(94.3, 96.9)}\tabularnewline
``dsm0111'' & 95.1 & \multicolumn{2}{c}{(93.8, 96.2)} & 95.8 & \multicolumn{2}{c}{(94.6, 97.2)}\tabularnewline
\hline 
\end{tabular}
\end{table}

\section{Real-data application\label{sec:Real-data-application}}

In this section, we apply the proposed DSM method as well as other
existing methods in Section \ref{sec:Simulation-study} to the well-known
National Supported Work (NSW) data \citep{lalonde1986evaluating,firpo2007efficient}.
This dataset documented the effect of a job training program for the
unemployed on future earnings. \textcolor{black}{Following \citet{dehejia1999causal}},
we include the comparison group from Westat's Matched Current Population
Survey-Social Security (CPS) Administration File. In our analysis,
we include $185$ treated units and $689$ control units from the
NSW, as well as $429$ comparison units from the CPS-3, a subset of
the CPS data \citep{lalonde1986evaluating,firpo2007efficient}. Seven
baseline confounding covariates are used for this application: age,
education, race, Hispanic, married, having no college degree, and
real earnings in 1975. The outcome of interest is the real earnings
in 1978.

Because the outcome distributions are highly skewed (see Figure \ref{fig2:lalonde}),
the average treatment effect may not provide a comprehensive evaluation
of the job training program. Therefore, we estimate the ATT and QTTs.
The propensity score is estimated based on a logistic regression model
with all first-order terms of the covariates\textcolor{black}{{} and
second-order terms of numerical variables, following \citet{dehejia1999causal}.
The prognostic score is estimated based on a linear regression of
the earnings with the same predictors as in the propensity score model
for the control group.}

\begin{table}
\centering \caption{\label{t:check_bal}Covariate balance check before and after DSM}
\begin{tabular}{ccccccccc}
\hline 
 &  & age & educ & black & hisp & married & nodegr & re75\tabularnewline
\hline 
Before & Treatment group mean & 24.63 & 10.38 & 0.80 & 0.09 & 0.17 & 0.73 & 3066\tabularnewline
Matching & Control group mean & 26.25 & 10.21 & 0.50 & 0.13 & 0.34 & 0.70 & 2745\tabularnewline
 & Stand diff. in means & -0.19 & 0.08 & 0.61 & -0.10 & -0.37 & 0.06 & 0.07\tabularnewline
\hline 
After & Treatment group mean & 24.63 & 10.38 & 0.80 & 0.09 & 0.17 & 0.73 & 3066\tabularnewline
Matching & Control group mean & 25.04 & 10.39 & 0.80 & 0.11 & 0.17 & 0.72 & 2898\tabularnewline
 & Stand diff. in means & -0.05 & 0.00 & 0.01 & -0.05 & -0.01 & 0.01 & 0.04\tabularnewline
\hline 
\end{tabular}
\end{table}
\begin{table}
\centering \caption{\label{t:res}Estimated ATT and QTTs at the $0.1$, $0.25$, $0.3$,
$0.5$, $0.75$ and $0.9$ quantiles, and $95\%$ Wald confidence
intervals}

\resizebox{\textwidth}{!}{
\begin{tabular}{cccccccc}
\hline 
Estimand & m.x & psm & pgm & dsm &  &  & Naive(ATE\&QTE)\tabularnewline
\hline 
ATT & 372 (-746,1489) & 918 (-222,2058) & -150 (-1215,914) & 1088 (-57,2233) &  &  & -65 (-957,827)\tabularnewline
0.1-QTT & 0 (0,0) & 0 (0,0) & 0 (0,0) & 0 (0,0) &  &  & 0 (0,0)\tabularnewline
0.25-QTT & 549 (-90,1189) & 549 (-55,1154) & 549 (-104,1202) & 549 (-76,1175) &  &  & 549 (-55,1153)\tabularnewline
0.3-QTT & 935 (57,1813) & 1064 (268,1860) & 1039 (96,1982) & 1019 (134,1904) &  &  & 604 (-641,1170)\tabularnewline
0.5-QTT & 524 (-1606,2654) & 889 (-741,2519) & 1296 (-252,2844) & 757 (-997,2511) &  &  & 69 (-1093,1230)\tabularnewline
0.75-QTT & -391 (-2752,1970) & 617 (-1451,2686) & 737 (-2070,3544) & 1763 (-656,4182) &  &  & -195 (-1441,1578)\tabularnewline
0.9-QTT & -963 (-3482,1556) & 897 (-1750,3543) & -936 (-4770,1795) & 864 (-1395,3123) &  &  & -2326 (-2021,2158)\tabularnewline
\hline 
\end{tabular}}
\end{table}

Matching admits a transparent assessment of covariate balance before
and after matching. Table \ref{t:check_bal} presents the means of
all covariates by treatment group and the standardized difference
in means before and after DSM. The standardized difference is calculated
as the difference of the group means divided by the overall standard
error in the original sample. DSM makes standardized differences fall
between $-0.05$ to $0.05$ for all covariates, reducing the differences
of the observed covariates in the treated and the control.

\begin{figure}
\centering \caption{\label{fig:Quantile-effect-plot}Quantile effect plot on the treated
from the DSM algorithm}

\includegraphics[scale=0.75]{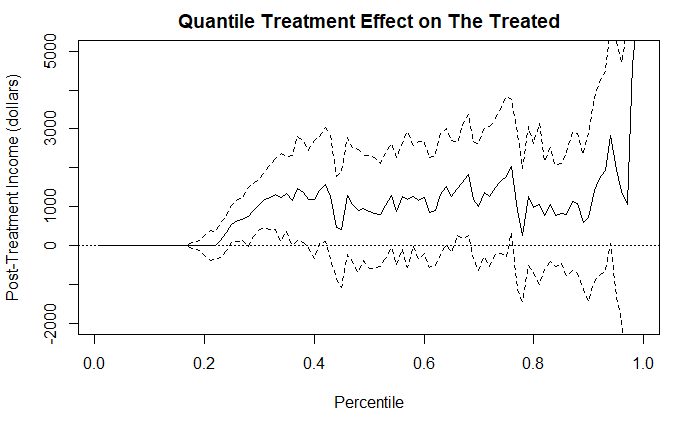}
\end{figure}

Table \ref{t:res} shows the estimated ATTs and QTTs at the $0.1$,
$0.25$, $0.3$, $0.5$, $0.75$ and $0.9$ quantiles, and $95\%$
Wald confidence intervals from the four matching methods, as well
as ATE and QTE estimated by naive method. All four matching estimators
show that the job training program does not have a significant effect
on the average earning for the treated. Figure \ref{fig:Quantile-effect-plot}
shows the QTT plot estimated by DSM algorithm. A closer inspection
of the QTT plot reveals that the effect is in fact significant around
percentile of $0.3$, which suggests that the program is beneficial
for the lower middle class.

\section{Discussion\label{sec:Discussion}}

We have developed multiply robust matching estimators for general
causal estimands. This framework offers a new ``metric'' to summarize
the differential roles of different covariates and also serves as
a powerful dimensional reduction tool in high-dimensional confounding.
The improved robustness comes from multiple model specifications for
the propensity score and prognostic score. The proposed DSM estimators
thus provide multiple protections to model misspecification and therefore
is an attractive alternative to existing weighting estimators.

Several issues are worth discussing. As with PSM, although the matching
variables are well balanced, individual covariates may not for a given
application. In this case, if the researchers know important confounders
based on substantive knowledge, they can augment the double score
by adding those confounders to ensure balance for these confounders;
however, adding too many variables will results in potential bias
as demonstrated in our simulation. Alternatively, one can use regression
adjustment for the matched sample \citet{abadie2016robust}, which
can remove remaining confounding biases. We focus on a binary treatment.
\citet{yang2016propensity} has developed the generalized propensity
score matching for estimating the treatment effects for more then
two treatments. Instead of creating a matched set to estimate the
treatment contrast directly, \citet{yang2016propensity} proposed
to create matched sets to estimate potential outcome means separately.
This approach allows matching based on one scalar function, namely
the generalized propensity score at a given treatment level, one at
a time. It is also of interest to extend our DSM algorithm to more
than two treatment comparison. It is important to highlight that as
for all existing matching methods, the DSM method cannot account for
unmeasured confounding. Following \citet{rosenbaum1983assessing}
and \citet{robins2000sensitivity}, we will develop sensitivity analyses
to no unmeasured confounding in the matching framework.

\section*{Acknowledgment}

We are grateful to Alberto Abadie for providing comments. Yang is
partially supported by the National Science Foundation grant DMS 1811245,
National Cancer Institute grant P01 CA142538, National Institute on
Aging grant 1R01AG066883, and National Institute of Environmental
Health Science grant 1R01ES031651. 

\bibliographystyle{dcu}
\bibliography{ci}

\pagebreak{}
\begin{center}
\textbf{\Large{}{}Supplementary Material for ``Multiply robust matching
estimators for average and quantile treatment effects''}{\Large\par}
\par\end{center}

\pagenumbering{arabic} 
\renewcommand*{\thepage}{S\arabic{page}}

\setcounter{lemma}{0} 
\global\long\def\thelemma{S\arabic{lemma}}%
\setcounter{equation}{0} 
\global\long\def\theequation{S\arabic{equation}}%
\setcounter{section}{0} 
\global\long\def\thesection{S\arabic{section}}%
\global\long\def\thesubsection{S\arabic{section}}%

\global\long\def\theassumption{S\arabic{assumption}}%
\setcounter{assumption}{0}

\global\long\def\thefigure{S\arabic{figure}}%
\setcounter{figure}{0}

Sections \ref{sec:Proof-of-Theorem1}, \ref{sec:Proof-for-Theorem2},
\ref{sec:Proof-of-Theorem 2}, and \ref{sec:Proof-of-Theorem3-q}
present the proofs of Theorems \ref{Thm:1}, \ref{Thm:3}, \ref{Thm:2},
and \ref{Thm:3-q}, respectively. Section \ref{sec:Comparison-of-PSM}
compares the efficiency of PSM, PGM, and DSM. Section \ref{sec:Le-Cam's-third}
presents Le Cam's third lemma. Section \ref{sec:Extensions-to-the}
presents the extensions to the ATT and QTT. Section \ref{sec:Figure}
presents a figure for the application. 

\section{Proof of Theorem \ref{Thm:1} \label{sec:Proof-of-Theorem1}}

For simplicity of the presentation, we omit the dependence of $\theta_{a}^{*}$
for $S_{a}(\theta_{a}^{*})$ if there is no ambiguity. Following \citet{abadie2011bias}
and \citet{abadie2012martingale}, under mild regularity conditions
on the nonparametric estimation, we have $\hat{B}_{n}=B_{n}+o_{P}(1)$.
Then, $\hat{\tau}_{\dsm}(\theta^{*})$ has the following asymptotic
linear form:
\begin{eqnarray}
n^{1/2}\left\{ \hat{\tau}_{\dsm}(\theta^{*})-\tau\right\}  & = & n^{-1/2}\sum_{i=1}^{n}\left\{ \mu_{1}(S_{1,i})-\mu_{0}(S_{0,i})-\tau\right\} \label{eq:A2-1}\\
 &  & +n^{-1/2}\sum_{i=1}^{n}A_{i}\left(1+M^{-1}K_{S_{1},i}\right)\left\{ Y_{i}-\mu_{1}(S_{1,i})\right\} \\
 &  & -n^{-1/2}\sum_{i=1}^{n}(1-A_{i})\left(1+M^{-1}K_{S_{0},i}\right)\left\{ Y_{i}-\mu_{0}(S_{0,i})\right\} +o_{P}(1).
\end{eqnarray}
If any model of the propensity score or prognostic score is correctly
specified, by Lemma \ref{lem3:newds}, we have $\E\{\mu_{1}(S_{1,i})-\mu_{0}(S_{0,i})\}=\tau$
and therefore $n^{1/2}\left\{ \hat{\tau}_{\dsm}(\theta^{*})-\tau\right\} $
converges to zero.

Let the three terms in (\ref{eq:A2-1}) be 
\begin{eqnarray*}
T_{1n} & = & n^{-1/2}\sum_{i=1}^{n}\left\{ \mu_{1}(S_{1,i})-\mu_{0}(S_{0,i})-\tau\right\} ,\\
T_{2n} & = & n^{-1/2}\sum_{i=1}^{n}A_{i}\left(1+M^{-1}K_{S_{1},i}\right)\left\{ Y_{i}-\mu_{1}(S_{1,i})\right\} ,\\
T_{3n} & = & -n^{-1/2}\sum_{i=1}^{n}(1-A_{i})\left(1+M^{-1}K_{S_{0},i}\right)\left\{ Y_{i}-\mu_{0}(S_{0,i})\right\} .
\end{eqnarray*}
We show the covariances of the three terms are zero: 
\begin{eqnarray*}
\cov(T_{1n},T_{2n}) & = & n^{-1}\sum_{i=1}^{n}\sum_{j=1}^{n}\cov\left[\mu_{1}(S_{1,i})-\mu_{0}(S_{0,i})-\tau,A_{j}\left(1+M^{-1}K_{S_{1},j}\right)\left\{ Y_{j}-\mu_{1}(S_{1,j})\right\} \right]\\
 & = & n^{-1}\sum_{i=1}^{n}\cov\left[\mu_{1}(S_{1,i})-\mu_{0}(S_{0,i})-\tau,A_{i}\left(1+M^{-1}K_{S_{1},i}\right)\left\{ Y_{i}-\mu_{1}(S_{1,i})\right\} \right]\\
 & = & n^{-1}\sum_{i=1}^{n}\cov\left(\E\left\{ \mu_{1}(S_{1,i})-\mu_{0}(S_{0,i})-\tau\mid S_{1,i},S_{0,i}\right\} ,\right.\\
 &  & \left.\E\left[A_{i}\left(1+M^{-1}K_{S_{1},i}\right)\left\{ Y_{i}-\mu_{1}(S_{1,i})\right\} \mid S_{1,i},S_{0,i}\right]\right)\\
 &  & +n^{-1}\sum_{i=1}^{n}\E\left(\cov\left\{ \mu_{1}(S_{1,i})-\mu_{0}(S_{0,i})-\tau\mid S_{1,i},S_{0,i},\right.\right.\\
 &  & \left.\left.A_{i}\left(1+M^{-1}K_{S_{1},i}\right)\left\{ Y_{i}-\mu_{1}(S_{1,i})\right\} \mid S_{1,i},S_{0,i}\right\} \right)\\
 & = & 0,
\end{eqnarray*}
similarly, $\cov(T_{1n},T_{3n})=0$, and by construction, $\cov(T_{2n},T_{3n})=0$.
Thus, the asymptotic variance of $n^{1/2}\left\{ \hat{\tau}_{\dsm}(\theta^{*})-\tau\right\} $
is
\begin{eqnarray*}
\V\left[n^{-1/2}\sum_{i=1}^{n}\left\{ \mu_{1}(S_{1,i})-\mu_{0}(S_{0,i})-\tau\right\} \right] & + & \V\left[n^{-1/2}\sum_{i=1}^{n}A_{i}\left(1+M^{-1}K_{S_{1},i}\right)\left\{ Y_{i}-\mu_{1}(S_{1,i})\right\} \right]\\
 & + & \V\left[n^{-1/2}\sum_{i=1}^{n}(1-A_{i})\left(1+M^{-1}K_{S_{0},i}\right)\left\{ Y_{i}-\mu_{0}(S_{0,i})\right\} \right].
\end{eqnarray*}
The first term becomes $\E\left[\{\mu_{1}(S_{1})-\mu_{0}(S_{0})-\tau\}^{2}\right].$
Following \citet{abadie2006large}, the second and third term, as
$n\rightarrow\infty$, becomes
\begin{eqnarray*}
 &  & \plim_{n\rightarrow\infty}\left[n^{-1}\sum_{i=1}^{n}A_{i}\left(1+M^{-1}K_{S_{1},i}\right)^{2}\V(Y_{i}\mid S_{1,i})\right]\\
 &  & +\plim_{n\rightarrow\infty}\left[n^{-1}\sum_{i=1}^{n}(1-A_{i})\left(1+M^{-1}K_{S_{0},i}\right)^{2}\V(Y_{i}\mid S_{0,i})\right]\\
 & = & \E\left(\sigma_{1}^{2}(S_{1})\left[\frac{1}{e(S_{1})}+\frac{1}{2M}\left\{ \frac{1}{e(S_{1})}-e(S_{1})\right\} \right]\right)\\
 &  & +\E\left(\sigma_{0}^{2}(S_{0})\left[\frac{1}{1-e(S_{0})}+\frac{1}{2M}\left\{ \frac{1}{1-e(S_{0})}-1+e(S_{0})\right\} \right]\right).
\end{eqnarray*}

\section{Comparison of PSM, PGM, and DSM\label{sec:Comparison-of-PSM}}

We compare the asymptotic variances of the PSM, PGM, and DSM estimators
of the ATE. To simplify the discussion, we assume one working model
$e(X;\alpha)$ for the propensity score and one working model $\Psi(X;\beta)=\{\Psi_{0}(X;\beta_{0}),\Psi_{1}(X;\beta_{1})\}$
for the prognostic score. In fact, the derivation for the DSM estimator
in Section \ref{sec:Proof-of-Theorem1} applies to the PSM estimator
and the PGM estimator by replacing $S_{a}=S_{a}(\theta_{a}^{*})=\{e(X;\alpha^{*}),\Psi_{a}(X;\beta_{a}^{*})\}$
($a=0,1$) with $e(X;\alpha^{*})$ and $\Psi_{a}(X;\beta^{*})$, respectively.

If \textit{the prognostic score model is correctly specified}, we
have $\Psi_{a}(X;\beta_{a}^{*})=\Psi_{a}(X)$ and therefore
\begin{equation}
\mu_{a}(S_{a})=\E\{Y(a)\mid S_{a}\}=\E\{Y(a)\mid e(X;\alpha^{*}),\Psi_{a}(X)\}=\E\{Y(a)\mid X\}=\mu_{a}(X),\label{eq:muaX}
\end{equation}
for $a=0,1$. Then, for the DSM estimator, $V_{\tau}$ in (\ref{eq:V1})
becomes
\begin{multline*}
V_{\tau,\dsm}=\E\left[\{\mu_{1}(X)-\mu_{0}(X)-\tau\}^{2}\right]+\E\left(\sigma_{1}^{2}(X)\left[\frac{1}{e(S_{1})}+\frac{1}{2M}\left\{ \frac{1}{e(S_{1})}-e(S_{1})\right\} \right]\right)\\
+\E\left(\sigma_{0}^{2}(X)\left[\frac{1}{1-e(S_{0})}+\frac{1}{2M}\left\{ \frac{1}{1-e(S_{0})}-1+e(S_{0})\right\} \right]\right).
\end{multline*}
For the PGM estimator, it is easy to derive that the corresponding
asymptotic variance is 
\begin{multline*}
V_{\tau,\prog}=\E\left[\{\mu_{1}(X)-\mu_{0}(X)-\tau\}^{2}\right]+\E\left\{ \sigma_{1}^{2}(X)\left(\frac{1}{e\{\Psi_{1}(X)\}}+\frac{1}{2M}\left[\frac{1}{e\{\Psi_{1}(X)\}}-e\{\Psi_{1}(X)\}\right]\right)\right\} \\
+\E\left\{ \sigma_{0}^{2}(X)\left(\frac{1}{1-e\{\Psi_{0}(X)\}}+\frac{1}{2M}\left[\frac{1}{1-e\{\Psi_{0}(X)\}}-1+e\{\Psi_{0}(X)\}\right]\right)\right\} .
\end{multline*}
By Jensen's inequality, we have
\begin{eqnarray*}
\E\left\{ \frac{1}{e(S_{a})}\mid\Psi_{a}(X)\right\}  & \geq & \frac{1}{\E\{e(S_{a})\mid\Psi_{a}(X)\}}=\frac{1}{e\{\Psi_{a}(X)\}},\\
\E\left\{ \frac{1}{1-e(S_{a})}\mid\Psi_{a}(X)\right\}  & \geq & \frac{1}{\E\{1-e(S_{a})\mid\Psi_{a}(X)\}}=\frac{1}{1-e\{\Psi_{a}(X)\}},
\end{eqnarray*}
for $a=0,1$. It follows that $V_{\tau,\dsm}\geq V_{\tau,\prog}$.

If \textit{the propensity score model is correctly specified}, we
have $e(X;\alpha^{*})=e(X)$ and therefore
\begin{equation}
e(S_{a})=\bP(A=1\mid S_{a})=\bP\{A=1\mid e(X),\Psi_{a}(X;\beta_{a}^{*})\}=e(X).\label{eS}
\end{equation}
Then, for the DSM estimator, $V_{\tau}$ in (\ref{eq:V1}) becomes
\begin{multline*}
V_{\tau,\dsm}=\E\left[\{\mu_{1}(S_{1})-\mu_{0}(S_{0})-\tau\}^{2}\right]+\E\left(\sigma_{1}^{2}(S_{1})\left[\frac{1}{e(X)}+\frac{1}{2M}\left\{ \frac{1}{e(X)}-e(X)\right\} \right]\right)\\
+\E\left(\sigma_{0}^{2}(S_{0})\left[\frac{1}{1-e(X)}+\frac{1}{2M}\left\{ \frac{1}{1-e(X)}-1+e(X)\right\} \right]\right).
\end{multline*}
For the PSM estimator, it is easy to derive that the corresponding
asymptotic variance is 
\begin{multline*}
V_{\tau,\psm}=\E\left(\left[\mu_{1}\{e(X)\}-\mu_{0}\{e(X)\}-\tau\right]{}^{2}\right)+\E\left(\sigma_{1}^{2}\{e(X)\}\left[\frac{1}{e(X)}+\frac{1}{2M}\left\{ \frac{1}{e(X)}-e(X)\right\} \right]\right)\\
+\E\left(\sigma_{0}^{2}\{e(X)\}\left[\frac{1}{1-e(X)}+\frac{1}{2M}\left\{ \frac{1}{1-e(X)}-1+e(X)\right\} \right]\right).
\end{multline*}
To compare $V_{\tau,\dsm}$ and $V_{\tau,\psm}$, we decompose 
\[
Y(a)=\underbrace{\mu_{a}\{e(X)\}+\epsilon_{a,S_{a}\mid e(X)}}_{\mu_{a}(S_{a})}+\epsilon_{a},
\]
where $\epsilon_{a,S_{a}\mid e(X)}$ and $\epsilon_{a}$ have mean
zero and satisfy that $\mu_{a}\{e(X)\}\indep\{\epsilon_{a,S_{a}\mid e(X)},\epsilon_{a}:a=0,1\}$
and $\epsilon_{a,S_{a}\mid e(X)}\indep\epsilon_{a}$. With this decomposition,
$\sigma_{a}^{2}\{e(X)\}=\E\{\epsilon_{a,S_{a}\mid e(X)}^{2}\mid e(X)\}+\E(\epsilon_{a}^{2})$
and $\sigma_{a}^{2}(S_{a})=\E(\epsilon_{a}^{2})$. Then, it follows
that 
\begin{eqnarray*}
V_{\tau,\psm}-V_{\tau,\dsm} & = & -\E\left\{ \left(\epsilon_{1,S_{1}\mid e(X)}-\epsilon_{0,S_{0}\mid e(X)}\right)^{2}\right\} +\E\left(\epsilon_{1,S_{1}\mid e(X)}^{2}\left[\frac{1}{e(X)}+\frac{1}{2M}\left\{ \frac{1}{e(X)}-e(X)\right\} \right]\right)\\
 &  & +\E\left(\epsilon_{0,S_{0}\mid e(X)}^{2}\left[\frac{1}{1-e(X)}+\frac{1}{2M}\left\{ \frac{1}{1-e(X)}-1+e(X)\right\} \right]\right)\\
 & = & 2\E\left\{ \epsilon_{1,S_{1}\mid e(X)}\epsilon_{0,S_{0}\mid e(X)}\right\} +\E\left(\epsilon_{1,S\mid e(X)}^{2}\left[\frac{1}{e(X)}-1+\frac{1}{2M}\left\{ \frac{1}{e(X)}-e(X)\right\} \right]\right)\\
 &  & +\E\left(\epsilon_{0,S_{0}\mid e(X)}^{2}\left[\frac{1}{1-e(X)}-1+\frac{1}{2M}\left\{ \frac{1}{1-e(X)}-1+e(X)\right\} \right]\right).
\end{eqnarray*}
The last two terms are always non-negative; however, the sign of $2\E\left\{ \epsilon_{1,S_{1}\mid e(X)}\epsilon_{0,S_{0}\mid e(X)}\right\} $
and therefore that of $V_{\tau,\psm}-V_{\tau,\dsm}$ can be either
positive, negative, or zero. Therefore, for estimating $\tau$, it
is not guaranteed that DSM is more efficient than PSM. For estimating
$\mu_{a}$, using the similar argument as above, $2\E\left\{ \epsilon_{1,S_{1}\mid e(X)}\epsilon_{0,S_{0}\mid e(X)}\right\} $
is absent, so DSM is more efficient than PSM. 

\section{Le Cam's third Lemma\label{sec:Le-Cam's-third}}

Consider two sequences of probability measures $(\mathbb{Q}^{(n)})_{n=1}^{\infty}$
and $(\mathbb{P}^{(n)})_{n=1}^{\infty}$. Assume that under $\mathbb{P}^{(n)}$,
a statistic $T_{n}$ and the likelihood ratios $\de\mathbb{Q}^{(n)}/\de\mathbb{P}^{(n)}$
satisfy 
\[
\left(\begin{array}{c}
T_{n}\\
\log(\de\mathbb{Q}^{(n)}/\de\mathbb{P}^{(n)})
\end{array}\right)\rightarrow\N\left\{ \left(\begin{array}{c}
0\\
-\sigma^{2}/2
\end{array}\right),\left(\begin{array}{cc}
\tau^{2} & c\\
c & \sigma^{2}
\end{array}\right)\right\} 
\]
in distribution, as $n\rightarrow\infty$. Then, under $\mathbb{Q}^{(n)}$,
\[
T_{n}\rightarrow\N(c,\tau^{2})
\]
in distribution, as $n\rightarrow\infty$. See \citet{le1990asymptotics},
\citet{bickel1993efficient}, and \citet{van1998asymptotic} for textbook
discussions.

\section{Proof of Theorem \ref{Thm:3} \label{sec:Proof-for-Theorem2} }

We follow the technique in \citet{andreou2012alternative} and \citet{abadie2016matching}.
In \citet{abadie2016matching}, the PSM estimators rely on the nuisance
parameter estimator under a correct specification of the propensity
score model. In our setting, the nuisance parameters include both
parameters in the propensity score model and the prognostic score
model, and require only one of the models to be correctly specified.
Without loss of generality, we assume one working model $e(X;\alpha)$
for the propensity score and one working model $\Psi(X;\beta)=\{\Psi_{0}(X;\beta_{0}),\Psi_{1}(X;\beta_{1})\}$
for the prognostic score. The proof for the case with more than two
working models for each score is similar at the expense of heavier
notation. Let\textcolor{blue}{{} }\textcolor{black}{$\bP$ be the distribution
of $\{(A_{i},X_{i},Y_{i}):$ $i=1,\ldots,n\}$}. Consider $\bP=\bP^{\theta^{*}}$
to be indexed by $\theta^{*}=(\alpha^{*\T},\beta_{0}^{*\T},\beta_{1}^{*\T})^{\T}$,
which satisfies
\begin{equation}
\E\{U(A,X,Y;\theta^{*})\}=\E\left\{ \left(\begin{array}{c}
U_{1}(A,X;\alpha^{*})\\
U_{2}(A,X,Y;\beta_{0}^{*})\\
U_{3}(A,X,Y;\beta_{1}^{*})
\end{array}\right)\right\} =0.\label{eq:ee}
\end{equation}

We invoke standard regularity conditions on Z-estimation \citep{van1998asymptotic}
as follows.

\begin{assumption}\label{asump:reg} (i) Under $\bP^{\theta^{*}}$,
$\cU_{n}(\theta^{*})\rightarrow\N(0,\Sigma_{U})$ in distribution,
as $n\rightarrow\infty$, where $\Sigma_{U}=\E\{U(A,X,Y;\theta^{*})$
$U(A,X,Y;\theta^{*})^{\T}\}$; (ii) $\Gamma_{\theta}=\E\{\partial U(A,X,Y;\theta)/\partial\theta^{\T}\}$
is nonsingular around $\theta^{*}$; and (iii) for any vector of constant
$h$, $\exp\{n^{1/2}h^{\T}\Gamma_{\theta^{*}}\Sigma_{U}^{-1}\cU_{n}(\theta^{*})\}$
is uniformly integrable.

\end{assumption} 

Under Assumption \ref{asump:reg},
\begin{equation}
n^{1/2}(\hat{\theta}-\theta^{*})=-\Gamma_{\theta^{*}}^{-1}\cU_{n}(\theta^{*})+o_{P}(1)\rightarrow\N(0,\Sigma_{\theta^{*}}),\label{eq:mle}
\end{equation}
 in distribution, as $n\rightarrow\infty$, where $\Sigma_{\theta^{*}}=\Gamma_{\theta^{*}}^{-1}\Sigma_{U}(\Gamma_{\theta^{*}}^{-1})^{\T}$.

To derive the large sample distribution of $\hat{\tau}_{\dsm}(\hat{\theta})$,
following \citet{abadie2016matching}, we impose the following regularity
conditions.

\begin{assumption}\label{asmp:tau-1}

There exists a neighborhood of $\theta^{*},$ such that for any $\theta$
in this region, the following conditions hold: for $a=0,1$, (i) the
matching variable $S_{a}(\theta)$ has a compact and convex support
$\mathcal{S}$, with a continuous density bounded and bounded away
from zero; (ii) $\mu_{a}\{S_{a}(\theta)\}$ and \textbf{$\sigma_{a}^{2}\{S_{a}(\theta)\}$
}satisfy the Lipschitz continuity condition; and (iii) there exists
$\delta>0$ such that $\E\left\{ |Y(a)|^{2+\delta}\mid S_{a}(\theta)\right\} $
is uniformly bounded for any $S_{a}(\theta)\in\mathcal{S}.$\end{assumption}

Following \citet{andreou2012alternative}, because we consider a semiparametric
model for $\theta^{*}$, to invoke the Le Cam's lemma, we specify
an auxiliary parametric model $\bP^{\theta_{n}}$ defined locally
though $\theta^{*}$, $\theta_{n}=\theta^{*}+n^{-1/2}h$, with a density
\begin{equation}
\frac{\exp\left\{ n^{1/2}(\theta_{n}-\theta^{*})^{\T}\Gamma_{\theta^{*}}\Sigma_{U}^{-1}\cU_{n}(\theta^{*})-2^{-1}n(\theta_{n}-\theta^{*})^{\T}\Sigma_{\theta^{*}}^{-1}(\theta_{n}-\theta^{*})\right\} }{\E\left[\exp\left\{ n^{1/2}(\theta_{n}-\theta^{*})^{\T}\Gamma_{\theta^{*}}\Sigma_{U}^{-1}\cU_{n}(\theta^{*})-2^{-1}n(\theta_{n}-\theta^{*})^{\T}\Sigma_{\theta^{*}}^{-1}(\theta_{n}-\theta^{*})\right\} \right]}.\label{eq:par}
\end{equation}
By Assumption \ref{asump:reg} (iii), $\exp\{n^{1/2}(\theta_{n}-\theta^{*})^{\T}\Gamma_{\theta^{*}}\Sigma_{U}^{-1}\cU_{n}(\theta^{*})\}$
is uniformly integrable, and thus model (\ref{eq:par}) is uniformly
locally asymptotically normal. Because under $\bP^{\theta^{*}}$,
$\cU_{n}(\theta^{*})\rightarrow\N(0,\Sigma_{U})$ in distribution,
the normalizing constant in the denominator converges to one as $n\rightarrow\infty$.
The Fisher information under the parametric model (\ref{eq:par})
is $n\Sigma_{\theta^{*}}^{-1}.$ Therefore, $\hat{\theta}$ is efficient
under model (\ref{eq:par}).

\textcolor{black}{Now consider $(A_{i},X_{i},Y_{i})$, for $i=1,\ldots,n$,}
with the local shift $\bP^{\theta_{n}}$ (\citealp{bickel1993efficient}).
Under model (\ref{eq:par}), the likelihood ratio under $\bP^{\theta_{n}}$
is
\begin{eqnarray}
\log(\de\bP^{\theta^{*}}/\de\bP^{\theta_{n}}) & = & -h^{\T}\Gamma_{\theta^{*}}\Sigma_{U}^{-1}\cU_{n}(\theta^{*})+\frac{1}{2}h^{\T}\Sigma_{\theta^{*}}^{-1}h+o_{p}(1)\nonumber \\
 & = & -h^{\T}\Gamma_{\theta^{*}}\Sigma_{U}^{-1}\cU_{n}(\theta_{n})-\frac{1}{2}h^{\T}\Sigma_{\theta^{*}}^{-1}h+o_{p}(1),\label{eq:1}
\end{eqnarray}
where the second equality follows by the Taylor expansion of $\cU_{n}(\theta^{*})$
at $\theta_{n}$. Moreover, under $\bP^{\theta_{n}}$: $\cU_{n}(\theta_{n})\rightarrow\N(0,\Sigma_{U})$
in distribution, as $n\rightarrow\infty$, and 
\begin{equation}
n^{1/2}(\hat{\theta}-\theta_{n})=\Gamma_{\theta^{*}}^{-1}\cU_{n}(\theta_{n})+o_{P}(1).\label{eq:2}
\end{equation}

We also assume the following regularity condition.

\begin{assumption}\label{asump:leCam3rd} For all bounded continuous
functions $h(A,X,Y)$, the conditional expectation $\E_{\theta_{n}}\{h(A,X,Y)\}$
converges in distribution to $\E\{h(A,X,Y)$$\}$, where $\E_{\theta_{n}}(\cdot)$
is the expectation taken with respect to $P^{\theta_{n}}$.

\end{assumption}

We derive the results in Theorem \ref{Thm:3} in two steps.

In the first step, under $\bP^{\theta_{n}}${\small{},} we write $\tau=\tau(\theta_{n})$
to reflect its dependence on $\theta_{n}$; to be specific, we have
\[
\tau(\theta_{n})=\E\left[\mu_{1}\{S_{1}(\theta_{n})\}-\mu_{0}\{S_{0}(\theta_{n})\}\right].
\]
 We derive that under $\bP^{\theta_{n}}${\small{},}
\begin{equation}
\left(\begin{array}{c}
n^{1/2}\{\hat{\tau}_{\dsm}(\theta_{n})-\tau(\theta_{n})\}\\
n^{1/2}(\hat{\theta}-\theta_{n})\\
\log(\de\bP^{\theta^{*}}/\de\bP^{\theta_{n}})
\end{array}\right)\rightarrow\N\left\{ \left(\begin{array}{c}
0\\
0\\
-\frac{1}{2}h^{\T}\Sigma_{\theta^{*}}^{-1}h
\end{array}\right),\left(\begin{array}{ccc}
V_{\tau} & \gamma_{1}^{\T}\Gamma_{\theta^{*}}^{-1} & -\gamma_{1}^{\T}\Sigma_{U}^{-1}\Gamma_{\theta^{*}}h\\
\Gamma_{\theta^{*}}^{-1}\gamma_{1} & \Sigma_{\theta^{*}} & -h\\
-h^{\T}\Gamma_{\theta^{*}}\Sigma_{U}^{-1}\gamma_{1} & -h^{\T} & h^{\T}\Sigma_{\theta^{*}}^{-1}h
\end{array}\right)\right\} \label{eq:(11)}
\end{equation}
in distribution, as $n\rightarrow\infty$. We then express $\tau(\theta_{n})=\tau(\theta^{*})+\gamma_{2}^{\T}(n^{-1/2}h)+o(n^{-1/2})$,
where 
\begin{equation}
\gamma_{2}=\left.\frac{\partial\tau(\theta)}{\partial\theta}\right\vert _{\theta=\theta^{*}}=\E\left[\left.\frac{\partial\mu_{1}\{S_{1}(\theta)\}-\mu_{0}\{S_{0}(\theta)\}}{\partial\theta}\right\vert _{\theta=\theta^{*}}\right].\label{eq:gamma2}
\end{equation}
By Le Cam's third lemma, under $\bP^{\theta^{*}}$, 
\[
\left(\begin{array}{c}
n^{1/2}\{\hat{\tau}_{\dsm}(\theta_{n})-\tau\}\\
n^{1/2}(\hat{\theta}-\theta_{n})
\end{array}\right)\rightarrow\N\left\{ \left(\begin{array}{c}
-\gamma_{1}^{\T}\Sigma_{U}^{-1}\Gamma_{\theta^{*}}h-\gamma_{2}^{\T}h\\
-h
\end{array}\right),\left(\begin{array}{cc}
V_{\tau} & \gamma_{1}^{\T}\Gamma_{\theta^{*}}^{-1}\\
\Gamma_{\theta^{*}}^{-1}\gamma_{1} & \Sigma_{\theta^{*}}
\end{array}\right)\right\} 
\]
in distribution, as $n\rightarrow\infty$. Replacing $\theta_{n}$
by $\theta^{*}+n^{-1/2}h$ yields that under $\bP^{\theta^{*}}$\textcolor{black}{\small{},
}\textcolor{black}{
\begin{equation}
\left(\begin{array}{c}
n^{1/2}\{\hat{\tau}_{\dsm}(\theta^{*}+n^{-1/2}h)-\tau\}\\
n^{1/2}(\hat{\theta}-\theta^{*})
\end{array}\right)\rightarrow\N\left\{ \left(\begin{array}{c}
-\gamma_{1}^{\T}\Sigma_{U}^{-1}\Gamma_{\theta^{*}}h-\gamma_{2}^{\T}h\\
0
\end{array}\right),\left(\begin{array}{cc}
V_{\tau} & \gamma_{1}^{\T}\Gamma_{\theta^{*}}^{-1}\\
\Gamma_{\theta^{*}}^{-1}\gamma_{1} & \Sigma_{\theta^{*}}
\end{array}\right)\right\} \label{eq:3}
\end{equation}
}in distribution, as $n\rightarrow\infty$.

In the second step, we provide a heuristic derivation for (\ref{eq:3})
to obtain the approximate distribution (\ref{eq:sig2_adj}). If the
Normal distribution were exact, then 
\begin{equation}
n^{1/2}\{\hat{\tau}_{\dsm}(\theta^{*}+n^{-1/2}h)-\tau\}\mid n^{1/2}(\hat{\theta}-\theta^{*})=h\sim\N\left(-\gamma_{2}^{\T}h,V_{\tau}-\gamma_{1}^{\T}\Sigma_{U}^{-1}\gamma_{1}\right).\label{eq:(12)}
\end{equation}
Given that $n^{1/2}(\hat{\theta}-\theta^{*})=h$, we have $\theta^{*}+n^{-1/2}h=\hat{\theta}$,
and hence $\hat{\tau}_{\dsm}(\theta^{*}+n^{-1/2}h)=\hat{\tau}_{\dsm}(\hat{\theta})$.
Marginalizing (\ref{eq:(12)}) over the asymptotic distribution of
$n^{1/2}(\hat{\theta}-\theta^{*})$, we derive (\ref{eq:sig2_adj}).
The formal technique to derive (\ref{eq:sig2_adj}) can be find in
\citet{andreou2012alternative} and \citet{abadie2016matching}. To
avoid repetition, we omit this step.

In the following, we provide the proof to (\ref{eq:(11)}) in the
first step of the proof. Asymptotic normality of $n^{1/2}\{\hat{\tau}_{\dsm}(\theta_{n})-\tau(\theta_{n})\}$
under $\bP^{\theta_{n}}$ follows from Theorem \ref{Thm:1} and the
uniform local asymptotic normality of model (\ref{eq:par}). Asymptotic
joint normality of $\log(\de\bP^{\theta^{*}}/\de\bP^{\theta_{n}})$
and $n^{1/2}(\hat{\theta}-\theta_{n})$ follows from (\ref{eq:1})
and (\ref{eq:2}). Also, $n^{1/2}\{\hat{\tau}_{\dsm}(\theta_{n})-\tau(\theta_{n})\}=D_{n}(\theta_{n})+o_{P}(1)$,
where

\begin{eqnarray*}
D_{n}(\theta_{n}) & = & n^{-1/2}\sum_{i=1}^{n}\left[\mu_{1}\{S_{1,i}(\theta_{n})\}-\mu_{0}\{S_{0,i}(\theta_{n})\}-\tau(\theta_{n})\right]\\
 &  & +n^{-1/2}\sum_{i=1}^{n}A_{i}\left\{ 1+M^{-1}K_{S_{1}(\theta_{n}),i}\right\} \left[Y_{i}-\mu_{1}\{S_{1,i}(\theta_{n})\}\right]\\
 &  & -n^{-1/2}\sum_{i=1}^{n}(1-A_{i})\left\{ 1+M^{-1}K_{S_{0}(\theta_{n}),i}\right\} \left[Y_{i}-\mu_{0}\{S_{0,i}(\theta_{n})\}\right]+o_{P}(1).
\end{eqnarray*}
Therefore, the remaining is to show that, under $\bP^{\theta_{n}}$:
\begin{equation}
\left(\begin{array}{c}
D_{n}(\theta_{n})\\
\cU_{n}(\theta_{n})
\end{array}\right)\rightarrow\N\left\{ \left(\begin{array}{c}
0\\
0
\end{array}\right),\left(\begin{array}{cc}
V_{\tau} & \gamma_{1}^{\T}\\
\gamma_{1} & \Sigma_{U}
\end{array}\right)\right\} \label{eq:dist1}
\end{equation}
in distribution, as $n\rightarrow\infty$. To prove (\ref{eq:dist1}),
consider the linear combination
\begin{eqnarray*}
T_{n} & = & c_{0}D_{n}(\theta_{n})+c^{\T}\cU_{n}(\theta_{n})\\
 & = & c_{0}n^{-1/2}\sum_{i=1}^{n}\left[\mu_{1}\{S_{1,i}(\theta_{n})\}-\mu_{0}\{S_{0,i}(\theta_{n})\}-\tau(\theta_{n})\right]\\
 &  & +c_{0}n^{-1/2}\sum_{i=1}^{n}(2A_{i}-1)\left\{ 1+M^{-1}K_{S_{A_{i}}(\theta_{n}),i}\right\} \left[Y_{i}-\mu_{A_{i}}\{S_{A_{i},i}(\theta_{n})\}\right]\\
 &  & +c_{1}^{\T}n^{-1/2}\sum_{i=1}^{n}\left(\frac{\partial e(X_{i};\alpha_{n})}{\partial\alpha}\frac{A_{i}-e(X_{i};\alpha_{n})}{e(X_{i};\alpha_{n})\{1-e(X_{i};\alpha_{n})\}}\right)\\
 &  & +c_{2}^{\T}n^{-1/2}\sum_{i=1}^{n}\left((1-A_{i})\frac{\partial\mu_{0}(X_{i};\beta_{0,n})}{\partial\beta_{0}}\{Y_{i}-\mu_{0}(X_{i};\beta_{0,n})\}\right)\\
 &  & +c_{3}^{\T}n^{-1/2}\sum_{i=1}^{n}\left(A_{i}\frac{\partial\mu_{1}(X_{i};\beta_{1,n})}{\partial\beta_{1}}\{Y_{i}-\mu_{1}(X_{i};\beta_{1,n})\}\right)+o_{P}(1),
\end{eqnarray*}
where $c=(c_{1}^{\T},c_{2}^{\T},c_{3}^{\T})^{\T}$. We analyze $T_{n}$
using the martingale theory. We rewrite $T_{n}=\sum_{k=1}^{2n}\xi_{n,k},$
where 
\[
\xi_{n,k}=\begin{cases}
\sum_{j=1}^{8}\xi_{n,k}^{(j)}, & 1\leq k\leq n,\\
\sum_{j=9}^{11}\xi_{n,k}^{(j)}, & n+1\leq k\leq2n,
\end{cases}
\]
\begin{eqnarray*}
\xi_{n,k}^{(1)} & = & c_{0}n^{-1/2}\left[\mu_{1}\{S_{1,k}(\theta_{n})\}-\mu_{0}\{S_{0,k}(\theta_{n})\}-\tau(\theta_{n})\right],\\
\xi_{n,k}^{(2)} & = & c_{0}n^{-1/2}(2A_{k}-1)\left\{ 1+M^{-1}K_{S_{A_{k}}(\theta_{n}),k}\right\} \left[\mu_{A_{k}}(X_{k})-\mu_{A_{k}}\{S_{A_{k},k}(\theta_{n})\}\right]\\
\xi_{n,k}^{(3)} & = & c_{1}^{\T}n^{-1/2}\left(\frac{\partial e(X_{k};\alpha_{n})}{\partial\alpha}\frac{e(X_{k})-e(X_{k};\alpha_{n})}{e(X_{k};\alpha_{n})\{1-e(X_{k};\alpha_{n})\}}\right),\\
\xi_{n,k}^{(4)} & = & c_{2}^{\T}n^{-1/2}\{1-e(X_{k})\}\frac{\partial\mu_{0}(X_{k};\beta_{0,n})}{\partial\beta_{0}}\{\mu_{0}(X_{k})-\mu_{0}(X_{k};\beta_{0,n})\},\\
\xi_{n,k}^{(5)} & = & -c_{3}^{\T}n^{-1/2}e(X_{k})\frac{\partial\mu_{1}(X_{k};\beta_{1,n})}{\partial\beta_{1}}\{\mu_{1}(X_{k})-\mu_{1}(X_{k};\beta_{1,n})\},\\
\xi_{n,k}^{(6)} & = & c_{1}^{\T}n^{-1/2}\left(\frac{\partial e(X_{k};\alpha_{n})}{\partial\alpha}\frac{A_{k}-e(X_{k})}{e(X_{k};\alpha_{n})\{1-e(X_{k};\alpha_{n})\}}\right),\\
\xi_{n,k}^{(7)} & = & -c_{2}^{\T}n^{-1/2}\left(\{A_{k}-e(X_{k})\}\frac{\partial\mu_{0}(X_{k};\beta_{0,n})}{\partial\beta_{0}}\{\mu_{0}(X_{k})-\mu_{0}(X_{k};\beta_{0,n})\}\right),\\
\xi_{n,k}^{(8)} & = & c_{3}^{\T}n^{-1/2}\left(\{A_{k}-e(X_{k})\}\frac{\partial\mu_{1}(X_{k};\beta_{1,n})}{\partial\beta_{1}}\{\mu_{1}(X_{k})-\mu_{1}(X_{k};\beta_{1,n})\}\right),\\
\xi_{n,k}^{(9)} & = & c_{0}n^{-1/2}(2A_{k-n}-1)\left\{ 1+M^{-1}K_{S_{A_{k-n}}(\theta_{n}),k-n}\right\} \left\{ Y_{k-n}-\mu_{A_{k-n}}(X_{k-n})\right\} \\
\xi_{n,k}^{(10)} & = & c_{2}^{\T}n^{-1/2}(1-A_{k-n})\frac{\partial\mu_{0}(X_{k-n};\beta_{0,n})}{\partial\beta_{0}}\{Y_{k-n}-\mu_{0}(X_{k-n})\},\\
\xi_{n,k}^{(11)} & = & c_{3}^{\T}n^{-1/2}A_{k-n}\frac{\partial\mu_{1}(X_{k-n};\beta_{1,n})}{\partial\beta_{1}}\{Y_{k-n}-\mu_{1}(X_{k-n})\}.
\end{eqnarray*}
Consider the $\sigma$-fields
\[
\F_{n,k}=\begin{cases}
\sigma(A_{1},\ldots,A_{k},X_{1},\ldots,X_{k}), & 1\leq k\leq n,\\
\sigma(A_{1},\ldots,A_{n},X_{1},\ldots,X_{n},Y_{k-1},\ldots,Y_{k-n}), & 2n+1\leq k\leq3n.
\end{cases}
\]
Then, we have $\left\{ \sum_{k=1}^{i}\xi_{n,i},\F_{n,i},1\leq i\leq2n\right\} $
is a martingale for each $n\geq1$, which follows by the following
reasons:
\begin{description}
\item [{(i)}] because $S_{a,k}(\theta_{n})$ is a double balancing score,
\[
\E_{\theta_{n}}(\xi_{n,k}^{(1)}\mid\F_{n,k-1})=\E\left(c_{0}n^{-1/2}\left[\mu_{1}\{S_{1,k}(\theta_{n})\}-\mu_{0}\{S_{0,k}(\theta_{n})\}-\tau(\theta_{n})\right]\mid\F_{n,k-1}\right)=0;
\]
\item [{(ii)}] let $\F_{n,k}^{0}=\sigma\{A_{1},\ldots,A_{k},S_{1}(\theta_{n}),\ldots,S_{k}(\theta_{n})\}$
for $1\leq k\leq n$, then
\begin{eqnarray*}
\E_{\theta_{n}}(\xi_{n,k}^{(2)}\mid\F_{n,k-1}) & = & \E_{\theta_{n}}\{\E_{\theta_{n}}(\xi_{n,k}^{(2)}\mid\F_{n,k-1}^{0})\mid\F_{n,k-1}\}\\
 & = & c_{0}n^{-1/2}\E_{\theta_{n}}\left((2A_{k}-1)\left\{ 1+M^{-1}K_{S_{A_{k}}(\theta_{n}),k}\right\} \right.\\
 &  & \left.\times\E_{\theta_{n}}\left[\mu_{A_{k}}(X_{k})-\mu_{A_{k}}\{S_{A_{k},k}(\theta_{n})\}\mid\F_{n,k-1}^{0}\right]\mid\F_{n,k-1}\right)\\
 & = & c_{0}n^{-1/2}\E_{\theta_{n}}\left[(2A_{k}-1)\left\{ 1+M^{-1}K_{S_{A_{k}}(\theta_{n}),k}\right\} \times0\mid\F_{n,k-1}\right]\\
 & = & 0;
\end{eqnarray*}
\item [{(iii)}] $\mathcal{\E}_{\theta_{n}}(\xi_{n,k}^{(3)}\mid\F_{n,k-1})=\E_{\theta_{n}}(\xi_{n,k}^{(4)}\mid\F_{n,k-1})=\E_{\theta_{n}}(\xi_{n,k}^{(5)}\mid\F_{n,k-1})=0$
because $\E_{\theta_{n}}\{U(\theta_{n})\}=0$;
\item [{(iv)}] by the conditioning argument,
\[
\E_{\theta_{n}}(\xi_{n,k}^{(6)}\mid\F_{n,k-1})=\E_{\theta_{n}}\left[c_{1}^{\T}n^{-1/2}\frac{\partial e(X_{k};\alpha_{n})}{\partial\alpha}\frac{\E\left\{ A_{k}-e(X_{k})\mid\F_{n,k-1},X_{k}\right\} }{e(X_{k};\alpha_{n})\{1-e(X_{k};\alpha_{n})\}}\mid\F_{n,k-1}\right]=0;
\]
\item [{(v)}] $\E_{\theta_{n}}(\xi_{n,k}^{(7)}\mid\F_{n,k-1})=0$ and $\E_{\theta_{n}}(\xi_{n,k}^{(8)}\mid\F_{n,k-1})=0$
due to that fact that $A_{k}-e(X_{k})$ is unbiased conditional on
$X_{k}$;
\item [{(vi)}] $\E_{\theta_{n}}(\xi_{n,k}^{(9)}\mid\F_{n,k-1})=0$ because
$(1-A_{k-n})\{Y_{k-n}-\mu_{0}(X_{k-n})\}$ is unbiased given $\F_{n,k-1}$;
\item [{(vii)}] $\E_{\theta_{n}}(\xi_{n,k}^{(10)}\mid\F_{n,k-1})=0$ because
$A_{k-n}\{Y_{k-n}-\mu_{1}(X_{k-n})\}$ is unbiased given $\F_{n,k-1}$.
\end{description}
Therefore, we can apply the martingale central limit theorem \citep{billingsley1995probability}
to derive the limiting distribution of $T_{n}$. Under Assumption
\ref{asmp:tau-1}, we can verify the conditions for the martingale
central limit theorem hold. It follows that under $\bP^{\theta_{n}}$,
$T_{n}\rightarrow\N(0,\sigma^{2})$ in distribution, as $n\rightarrow\infty$,
where $\sigma^{2}=\plim\sum_{k=1}^{2n}\E_{\theta_{n}}(\xi_{n,k}^{2}\mid\F_{n,k-1})$.
Under Assumption \ref{asump:leCam3rd}, we thus derive the expression
of $\sigma^{2}$ and specify the components in (\ref{eq:dist1}) with
\begin{equation}
\gamma_{1}=(\gamma_{11}^{\T},\gamma_{12}^{\T},\gamma_{13}^{\T})^{\T},\label{eq:gamma1}
\end{equation}
\begin{eqnarray*}
\gamma_{11} & = & \E\left(\left[\mu_{1}\{S_{1}(\theta^{*})\}-\mu_{0}\{S_{0}(\theta^{*})\}-\tau\right]\frac{\partial e(X;\alpha^{*})}{\partial\alpha}\frac{A-e(X;\alpha^{*})}{e(X;\alpha^{*})\{1-e(X;\alpha^{*})\}}\right)\\
 &  & +\E\left(\left[\mu_{1}(X)-\mu_{1}\{S_{1}(\theta^{*})\}\right]\frac{\partial e(X;\alpha^{*})}{\partial\alpha}\frac{1-e(X;\alpha^{*})}{e(X;\alpha^{*})\{1-e(X;\alpha^{*})\}}\right)\\
 &  & -\E\left(\left[\mu_{0}(X)-\mu_{0}\{S_{0}(\theta^{*})\}\right]\frac{\partial e(X;\alpha^{*})}{\partial\alpha}\frac{-e(X;\alpha^{*})}{e(X;\alpha^{*})\{1-e(X;\alpha^{*})\}}\right),
\end{eqnarray*}
\begin{eqnarray*}
\gamma_{12} & = & -\E\left(\left[\mu_{1}\{S_{1}(\theta^{*})\}-\mu_{0}\{S_{0}(\theta^{*})\}-\tau\right](1-A)\frac{\partial\mu_{0}(X;\beta_{0}^{*})}{\partial\beta_{0}}\{\mu_{0}(X)-\mu_{0}(X;\beta_{0}^{*})\}\right)\\
 &  & -\E\left(\left[\mu_{0}(X)-\mu_{0}\{S_{0}(\theta^{*})\}\right]\frac{\partial\mu_{0}(X;\beta_{0}^{*})}{\partial\beta_{0}}\{\mu_{0}(X)-\mu_{0}(X;\beta_{0}^{*})\}\right)-\E\left\{ \frac{\partial\mu_{0}(X;\beta_{0}^{*})}{\partial\beta_{0}}\sigma_{0}^{2}(X)\right\} ,
\end{eqnarray*}
and
\begin{eqnarray*}
\gamma_{13} & = & -\E\left(\left[\mu_{1}\{S_{1}(\theta^{*})\}-\mu_{0}\{S_{0}(\theta^{*})\}-\tau\right]A\frac{\partial\mu_{1}(X;\beta_{1}^{*})}{\partial\beta_{1}}\{\mu_{1}(X)-\mu_{1}(X;\beta_{1}^{*})\}\right)\\
 &  & -\E\left(\left[\mu_{1}(X)-\mu_{1}\{S_{1}(\theta^{*})\}\right]\frac{\partial\mu_{1}(X;\beta_{1}^{*})}{\partial\beta_{1}}\{\mu_{1}(X)-\mu_{1}(X;\beta_{1}^{*})\}\right)-\E\left\{ \frac{\partial\mu_{1}(X;\beta_{1}^{*})}{\partial\beta_{1}}\sigma_{1}^{2}(X)\right\} .
\end{eqnarray*}

\section{Proof of Theorem \ref{Thm:2} \label{sec:Proof-of-Theorem 2}}

Under Assumption \ref{asmp:qte}, we can write 
\begin{equation}
\hat{F}_{a,\dsm}(\hat{q}_{a,\xi,\dsm})-F_{a}(q_{a,\xi})=\hat{F}_{a,\dsm}(q_{a,\xi})-F_{a}(q_{a,\xi})+f_{a}(q_{a,\xi})(\hat{q}_{a,\xi,\dsm}-q_{a,\xi})+o_{\mathrm{P}}(n^{-1/2}).\label{eq:s1}
\end{equation}
 Then the Bahadur-type representation for $\hat{q}_{a,\xi,\dsm}$
in (\ref{eq:quantile nni}) follows. 

We obtain the following decomposition
\begin{eqnarray*}
n^{1/2}\left\{ \hat{F}_{a,\dsm}^{(0)}(q)-F_{a}(q)\right\}  & = & n^{-1/2}\sum_{i=1}^{n}\left\{ \bone(A_{i}=a)\left(1+M^{-1}K_{S_{a},i}\right)\bone(Y_{i}\leq q)-F_{a}(q)\right\} \\
 & = & n^{-1/2}\sum_{i=1}^{n}\bone(A_{i}=a)\left(1+M^{-1}K_{S_{a},i}\right)\left\{ \bone(Y_{i}\leq q)-F_{a}(q;S_{a,i})\right\} \\
 &  & +n^{-1/2}\sum_{i=1}^{n}\left\{ \bone(A_{i}=a)\left(1+M^{-1}K_{S_{a},i}\right)F_{a}(q;S_{a,i})-F_{a}(q)\right\} \\
 & = & n^{-1/2}\sum_{i=1}^{n}\left\{ F_{a}(q;S_{a,i})-F_{a}(q)\right\} \\
 &  & +n^{-1/2}\sum_{i=1}^{n}\bone(A_{i}=a)\left(1+M^{-1}K_{S_{a},i}\right)\left\{ \bone(Y_{i}\leq q)-F_{a}(q;S_{a,i})\right\} \\
 &  & -n^{-1/2}\sum_{i=1}^{n}\bone(A_{i}=1-a)M^{-1}\sum_{j\in\J_{S_{a},i}}\left\{ F_{a}(q;S_{a,i})-F_{a}(q;S_{a,j})\right\} 
\end{eqnarray*}
and denote
\begin{eqnarray}
C_{a,n}(q) & = & n^{-1/2}\sum_{i=1}^{n}\left\{ F_{a}(q;S_{a,i})-F_{a}(q)\right\} \nonumber \\
 &  & +n^{-1/2}\sum_{i=1}^{n}\bone(A_{i}=a)\left(1+M^{-1}K_{S_{a},i}\right)\left\{ \bone(Y_{i}\leq q)-F_{a}(q;S_{a,i})\right\} ,\nonumber \\
B_{a,n}(q) & = & -n^{-1/2}\sum_{i=1}^{n}\bone(A_{i}=1-a)M^{-1}\sum_{j\in\J_{S_{a},i}}\left\{ F_{a}(q;S_{a,i})-F_{a}(q;S_{a,j})\right\} .\label{eq:B2_f}
\end{eqnarray}
Because of (\ref{eq:ps-balance}) and (\ref{eq:prog-balance}), $\E\{F_{a}(q;S_{a,i})\}=F_{a}(q)$,
so $\E\{C_{a,n}(q)\}=0$. The difference $F_{a}(q;S_{a,i})-F_{a}(q;S_{a,j})$
in (\ref{eq:B2_f}) accounts for the matching discrepancy, and therefore
$B_{a,n}(q)$ contributes to the asymptotic bias of the matching estimator.
To correct for the bias due to matching discrepancy, let $\hat{F}_{a}(q;S)$
be a nonparametric estimator of $F_{a}(q;S)$, for $a=0,1$. We propose
a de-biasing DSM estimator $\hat{F}_{a,\dsm}(q)$ of $F_{a}(q)$ in
(\ref{eq:F_a,dsm}). 

Therefore, by the Bahadur-type representation for $\hat{q}_{a,\xi,\dsm}$,
we have
\begin{eqnarray*}
n^{1/2}\left\{ \hat{\Delta}_{\xi,\dsm}(\theta^{*})-\Delta_{\xi}\right\}  & = & n^{1/2}\left\{ -\frac{\hat{F}_{1,\dsm}(q_{1,\xi})}{f_{1}(q_{1,\xi})}+\frac{\hat{F}_{0,\dsm}(q_{0,\xi})}{f_{0}(q_{0,\xi})}-\Delta_{\xi}\right\} +o_{P}(1)\\
 & = & n^{-1/2}\sum_{i=1}^{n}\left\{ -\frac{F_{1}(q_{1,\xi};S_{1,i})}{f_{1}(q_{1,\xi})}+\frac{F_{0}(q_{0,\xi};S_{0,i})}{f_{0}(q_{0,\xi})}-\Delta_{\xi}\right\} \\
 &  & -n^{-1/2}\sum_{i=1}^{n}(2A_{i}-1)\left\{ 1+M^{-1}K_{S_{A_{i}},i}\right\} \left\{ \frac{\bone(Y_{i}\leq q_{A_{i},\xi})-F_{A_{i}}(q_{A_{i},\xi};S_{A_{i},i})}{f_{A_{i}}(q_{A_{i},\xi})}\right\} +o_{P}(1).
\end{eqnarray*}
Following a similar derivation in the proof for Theorem \ref{Thm:1},
the asymptotic variance of $n^{1/2}\left\{ \hat{\Delta}_{\xi,\dsm}(\theta^{*})-\Delta_{\xi}\right\} $
is given by
\begin{multline}
V_{\Delta}=\E\left[\left\{ \frac{F_{1}(q_{1,\xi};S_{1})}{f_{1}(q_{1,\xi})}-\frac{F_{0}(q_{0,\xi};S_{0})}{f_{0}(q_{0,\xi})}\right\} ^{2}\right]\\
+\{f_{1}(q_{1,\xi})\}^{-2}\E\left(F_{1}(q_{1,\xi};S_{1})\{1-F_{1}(q_{1,\xi};S_{1})\}\left[\frac{1}{e(S_{1})}+\frac{1}{2M}\left\{ \frac{1}{e(S_{1})}-e(S_{1})\right\} \right]\right)\\
+\{f_{0}(q_{0,\xi})\}^{-2}\E\left(F_{0}(q_{0,\xi};S_{0})\{1-F_{0}(q_{0,\xi};S_{0})\}\left[\frac{1}{1-e(S_{0})}+\frac{1}{2M}\left\{ \frac{1}{1-e(S_{0})}-1+e(S_{0})\right\} \right]\right).\label{eq:V2}
\end{multline}

\section{Proof of Theorem \ref{Thm:3-q}\label{sec:Proof-of-Theorem3-q}}

The proof of Theorem \ref{Thm:3-q} is similar to that of Theorem
\ref{Thm:3}. It suffices to replace $n^{1/2}\{\hat{\tau}_{\dsm}(\theta_{n})-\tau\}$
in (\ref{eq:3}) by $n^{1/2}\{\hat{\Delta}_{\xi,\dsm}(\theta_{n})-\Delta_{\xi}\}$
and derive the corresponding $\gamma_{1}$ and $\gamma_{2}$, denoted
by $\gamma_{3}$ and $\gamma_{4}$. 

Toward that end, the key is to write
\begin{eqnarray*}
\hat{q}_{1,\xi,\dsm}-\hat{q}_{0,\xi,\dsm}-\Delta_{\xi}(\theta^{*}) & = & -\frac{\hat{F}_{1,\dsm}(q_{1,\xi})-F_{1}(q_{1,\xi})}{f_{1}(q_{1,\xi})}+\frac{\hat{F}_{0,\dsm}(q_{0,\xi})-F_{0}(q_{0,\xi})}{f_{0}(q_{0,\xi})}+o_{P}(n^{-1/2}),
\end{eqnarray*}
where
\begin{eqnarray*}
\hat{F}_{a,\dsm}(q_{a,\xi}) & = & n^{-1}\sum_{i=1}^{n}\left[F_{a}\{q_{a,\xi};S_{a}(\theta^{*})\}-\xi\right]\\
 &  & +n^{-1}\sum_{i=1}^{n}\bone(A_{i}=a)\left(1+M^{-1}K_{S_{a},i}\right)\left[\bone(Y_{i}\leq q_{a,\xi})-F_{a}\{q_{a,\xi};S_{a}(\theta^{*})\}\right],\\
\Delta_{\xi}(\theta^{*}) & = & \E\left[\frac{A}{e(X;\alpha^{*})}\frac{\bone(Y\leq q_{1,\xi})-F_{1}\{q_{1,\xi};S_{1}(\theta^{*})\}}{f_{1}(q_{1,\xi})}-\frac{1-A}{1-e(X;\alpha)}\frac{\bone(Y\leq q_{0,\xi})-F_{0}\{q_{0,\xi};S_{0}(\theta^{*})\}}{f_{0}(q_{0,\xi})}\right]\\
 &  & +\E\left[\frac{F_{1}\{q_{1,\xi};S_{1}(\theta^{*})\}}{f_{1}(q_{1,\xi})}-\frac{F_{0}\{q_{0,\xi};S_{0}(\theta^{*})\}}{f_{0}(q_{0,\xi})}\right],
\end{eqnarray*}
and repeat a similar analysis in Section \ref{sec:Proof-for-Theorem2}
with the following changes to $\xi_{n,k}^{(1)}$, $\xi_{n,k}^{(2)}$
and $\xi_{n,k}^{(9)}$: 
\begin{eqnarray*}
\tilde{\xi}_{n,k}^{(1)} & = & c_{0}n^{-1/2}\left[f_{1}(q_{1,\xi})^{-1}F_{1}\{q_{1,\xi};S_{1,k}(\theta_{n})\}-f_{0}(q_{0,\xi})^{-1}F_{0}\{q_{0,\xi};S_{0,k}(\theta_{n})\}-\Delta_{\xi}(\theta_{n})\right],\\
\tilde{\xi}_{n,k}^{(2)} & = & c_{0}n^{-1/2}(2A_{k}-1)\left\{ 1+M^{-1}K_{S_{A_{k}}(\theta_{n}),k}\right\} f_{A_{k}}(q_{A_{k},\xi})^{-1}\left[F_{A_{k}}(q_{A_{k},\xi};X_{k})-F_{A_{k}}\{q_{A_{k},\xi};S_{A_{k},k}(\theta_{n})\}\right],\\
\tilde{\xi}_{n,k}^{(9)} & = & c_{0}n^{-1/2}(2A_{k-n}-1)\left\{ 1+M^{-1}K_{S_{A_{k}}(\theta_{n}),k-n}\right\} f_{A_{k-n}}(q_{A_{k-n},\xi})^{-1}\\
 &  & \times\left\{ \bone(Y_{k-n}\leq q_{A_{k-n},\xi})-F_{A_{k-n}}(q_{A_{k-n},\xi};X_{k-n})\right\} .
\end{eqnarray*}
Then, we can derive 
\begin{equation}
\gamma_{3}^{\T}=(\gamma_{31}^{\T},\gamma_{32}^{\T},\gamma_{33}^{\T}),\label{eq:gamma3}
\end{equation}
where
\begin{eqnarray*}
\gamma_{31} & = & \E\left(\left[\frac{F_{1}\{q_{1,\xi};S_{1}(\theta^{*})\}}{f_{1}(q_{1,\xi})}-\frac{F_{0}\{q_{0,\xi};S_{0}(\theta^{*})\}}{f_{0}(q_{0,\xi})}-\Delta_{\xi}\right]\frac{\partial e(X;\alpha^{*})}{\partial\alpha}\frac{A-e(X;\alpha^{*})}{e(X;\alpha^{*})\{1-e(X;\alpha^{*})\}}\right)\\
 &  & +\E\left(f_{1}(q_{1,\xi})^{-1}\left[F_{1}(q_{1,\xi};X)-F_{1}\{q_{1,\xi};S_{1}(\theta^{*})\}\right]\frac{\partial e(X;\alpha^{*})}{\partial\alpha}\frac{1-e(X;\alpha^{*})}{e(X;\alpha^{*})\{1-e(X;\alpha^{*})\}}\right)\\
 &  & -\E\left(f_{0}(q_{0,\xi})^{-1}\left[F_{0}(q_{0,\xi};X)-F_{0}\{q_{0,\xi};S_{0}(\theta^{*})\}\right]\frac{\partial e(X;\alpha^{*})}{\partial\alpha}\frac{-e(X;\alpha^{*})}{e(X;\alpha^{*})\{1-e(X;\alpha^{*})\}}\right),
\end{eqnarray*}
\begin{eqnarray*}
\gamma_{32} & = & -\E\left(\left[\frac{F_{1}\{q_{1,\xi};S_{1}(\theta^{*})\}}{f_{1}(q_{1,\xi})}-\frac{F_{0}\{q_{0,\xi};S_{0}(\theta^{*})\}}{f_{0}(q_{0,\xi})}-\Delta_{\xi}\right](1-A)\frac{\partial\mu_{0}(X;\beta_{0}^{*})}{\partial\beta_{0}}\{\mu_{0}(X)-\mu_{0}(X;\beta_{0}^{*})\}\right)\\
 &  & -\E\left(f_{0}(q_{0,\xi})^{-1}\left[F_{0}(q_{0,\xi};X)-F_{0}\{q_{0,\xi};S_{0}(\theta^{*})\}\right]\frac{\partial\mu_{0}(X;\beta_{0}^{*})}{\partial\beta_{0}}\{\mu_{0}(X)-\mu_{0}(X;\beta_{0}^{*})\}\right)\\
 &  & -\E\left\{ \frac{\partial\mu_{0}(X;\beta_{0}^{*})}{\partial\beta_{0}}f_{0}(q_{0,\xi})^{-1}\left[I\{Y(0)\leq q_{0,\xi}\}-F_{0}(q_{0,\xi};X)\right]\{Y(0)-\mu_{0}(X)\}\right\} ,
\end{eqnarray*}
\begin{eqnarray*}
\gamma_{33} & = & -\E\left(\left[\frac{F_{1}\{q_{1,\xi};S_{1}(\theta^{*})\}}{f_{1}(q_{1,\xi})}-\frac{F_{0}\{q_{0,\xi};S_{0}(\theta^{*})\}}{f_{0}(q_{0,\xi})}-\Delta_{\xi}\right]A\frac{\partial\mu_{1}(X;\beta_{1}^{*})}{\partial\beta_{1}}\{\mu_{1}(X)-\mu_{1}(X;\beta_{1}^{*})\}\right)\\
 &  & -\E\left(f_{1}(q_{1,\xi})^{-1}\left[F_{1}(q_{1,\xi};X)-F_{1}\{q_{1,\xi};S_{1}(\theta^{*})\}\right]\frac{\partial\mu_{1}(X;\beta_{1}^{*})}{\partial\beta_{1}}\{\mu_{1}(X)-\mu_{1}(X;\beta_{1}^{*})\}\right)\\
 &  & -\E\left\{ \frac{\partial\mu_{1}(X;\beta_{1}^{*})}{\partial\beta_{1}}f_{1}(q_{1,\xi})^{-1}\left[\bone\{Y(1)\leq q_{1,\xi}\}-F_{1}(q_{1,\xi};X)\right]\{Y(1)-\mu_{1}(X)\}\right\} ,
\end{eqnarray*}
and 
\begin{equation}
\gamma_{4}=-\E\left[\frac{1}{f_{1}(q_{1,\xi})}\frac{\partial F_{1}\{q_{1,\xi};S_{1}(\theta^{*})\}}{\partial\theta}-\frac{1}{f_{0}(q_{0,\xi})}\frac{\partial F_{0}\{q_{0,\xi};S_{0}(\theta^{*})\}}{\partial\theta}\right].\label{eq:gamma4}
\end{equation}

\section{Extensions to the causal effects on the treated\label{sec:Extensions-to-the}}

In this extension, we estimate the average causal effect on the treated
$\tau_{\mathrm{ATT}}$ and the quantile treatment effect on the treated
$\Delta_{\mathrm{ATT},\xi}=q_{1,\xi\mid A=1}-q_{0,\xi\mid A=1}$,
where $q_{a,\xi\mid A=1}=\inf_{q}[P\{Y(a)\leq q\}\geq\xi\mid A=1]$,
$a=0,1$. Here, because $f\{Y(1)\mid A=1\}=f(Y\mid A=1)$, the outcome
distribution for the treated is identifiable. Therefore, $\E\{Y(1)\mid A=1\}=\E(Y\mid A=1)$
and $q_{1,\xi\mid A=1}=\inf_{q}\left\{ \bP(Y\leq q\mid A=1)\geq\xi\right\} $.

To identify the outcome distribution for the control, Assumptions
\ref{asump-ignorable} and \ref{asump-overlap} can be relaxed \citep{heckman1997matching}.

\begin{assumption} \label{asump-ignorable-1} $Y(0)\indep A\mid X$.
\end{assumption}

\begin{assumption} \label{asump-overlap-1}There exists a constant
$c$ such that $e(X)\leq c<1$ almost surely.

\end{assumption}

For the causal effects on the treated, the prognostic score $\Psi_{0}(X)$
is a sufficient statistic for $Y(0)$ in the sense that $Y(0)\indep X\mid\Psi_{0}(X)$
according to \citet{hansen2008prognostic}. Then, under Assumptions
\ref{asump-ignorable-1} and \ref{asump-overlap-1}, 
\begin{eqnarray*}
\tau_{\mathrm{ATT}} & = & \E[\E(Y\mid A=1)-\E\{Y\mid A=0,e(X)\}\mid A=1]\\
 & = & \E[\E(Y\mid A=1)-\E\{Y\mid A=0,\Psi_{0}(X)\}\mid A=1],
\end{eqnarray*}
and
\begin{eqnarray*}
q_{0,\xi\mid A=1} & = & \inf_{q}\left(\E[\bP\{Y\leq q\mid A=0,e(X)\}\mid A=1]\geq\xi\right)\\
 & = & \inf_{q}\left(\E[\bP\{Y\leq q\mid A=0,\Psi_{0}(X)\}\mid A=1]\geq\xi\right),
\end{eqnarray*}
encoding the double balancing properties of $S=\{e(X),\Psi_{0}(X)\}$.

The DSM estimators for $\tau_{\mathrm{ATT}}$ and $\Delta_{\mathrm{ATT},\xi}$
follow similar steps as for $\tau$ and $\Delta_{\xi}$. We describe
the differences below.

In the matching step, for each unit $i$ with treatment $A_{i}=1$,
find $M$ nearest neighbors from the control group $A_{i}=0$ based
on the matching variable $S_{i}=S_{i}(\hat{\theta})$. \textcolor{black}{Let
these matched units for unit $i$ be indexed by $\J_{S(\hat{\theta}),i}$.}

\textcolor{black}{The initial and de-biasing DSM estimators of $\tau_{\ATT}$
are}
\begin{eqnarray*}
\hat{\tau}_{\ATT,\dsm}^{(0)} & = & n_{1}^{-1}\sum_{i=1}^{n}A_{i}\{Y_{i}-\hat{Y}_{i}(0)\},\ \hat{Y}_{i}(0)=M^{-1}\sum_{j\in\J_{S,i}}Y_{j},\\
\hat{\tau}_{\ATT,\dsm} & = & \hat{\tau}_{\ATT,\dsm}^{(0)}-n_{1}^{-1}\sum_{i=1}^{n}A_{i}\left\{ \hat{\mu}_{0}(S_{i})-M^{-1}\sum_{j\in\J_{S,i}}\hat{\mu}_{0}(S_{j})\right\} .
\end{eqnarray*}

Let the estimator of $F_{1}(q\mid A=1)=\bP\{Y(1)<q\mid A=1\}$ be
\[
\hat{F}_{1}(q\mid A=1)=n_{1}^{-1}\sum_{i=1}^{n}A_{i}\bone(Y_{i}\leq q).
\]
Then, we estimate $q_{1,\xi\mid A=1}$ by 
\[
\hat{q}_{1,\xi\mid A=1}=\inf_{q}\{\hat{F}_{1}(q\mid A=1)\geq\xi\}.
\]
The \textcolor{black}{initial and de-biasing DSM estimators of $F_{0}(q\mid A=1)=\bP\{Y(0)<q\mid A=1\}$
are
\begin{eqnarray*}
\hat{F}_{0,\dsm}^{(0)}(q\mid A=1) & = & n_{1}^{-1}\sum_{i=1}^{n}A_{i}M^{-1}\sum_{j\in\J_{S,i}}\bone(Y_{j}\leq q)=n_{1}^{-1}\sum_{i=1}^{n}(1-A_{i})M^{-1}K_{S,i}\bone(Y_{i}\leq q),\\
\hat{F}_{0,\dsm}(q\mid A=1) & = & \hat{F}_{0,\dsm}^{(0)}(q\mid A=1)-n_{1}^{-1/2}\hat{B}_{0,n}(q),\\
\hat{B}_{0,n}(q) & = & -n_{1}^{-1/2}\sum_{i=1}^{n}A_{i}M^{-1}\sum_{j\in\J_{S,i}}\left\{ \hat{F}_{0}(q;S_{i})-\hat{F}_{0}(q;S_{j})\right\} .
\end{eqnarray*}
}Then, we estimate $q_{0,\xi\mid A=1}$ by 
\[
\hat{q}_{0,\xi\mid A=1,\dsm}=\inf_{q}\{\hat{F}_{0,\dsm}(q\mid A=1)\geq\xi\}.
\]
Lastly, the DSM estimator of $\Delta_{\ATT,\xi}$ is $\hat{\Delta}_{\ATT,\xi,\dsm}=\hat{q}_{1,\xi\mid A=1}-\hat{q}_{0,\xi\mid A=1,\dsm}$.

For variance estimation, we replace the VE-Step 2 and VE-Step 2' for
$\tau$ and $\Delta_{\xi}$ by the following steps:
\begin{description}
\item [{ATT-VE-Step$\ 2.$}] Obtain a bootstrap replicate of $\hat{\tau}_{\ATT,\dsm}(\hat{\theta})$,
\begin{multline*}
\hat{\tau}_{\ATT,\dsm}^{*}(\hat{\theta}^{*})=n_{1}^{-1}\sum_{i=1}^{n}\omega_{i}^{*}A_{i}\left[\hat{\mu}_{1}\{S_{i}(\hat{\theta}^{*})\}-\hat{\mu}_{0}\{S_{i}(\hat{\theta}^{*})\}\right]\\
+n_{1}^{-1}\sum_{i=1}^{n}\omega_{i}^{*}\left\{ A_{i}-(1-A_{i})M^{-1}K_{S(\hat{\theta}),i}\right\} \left[Y_{i}-\hat{\mu}_{A_{i}}\{S_{i}(\hat{\theta}^{*})\}\right].
\end{multline*}
\item [{QTT-VE-Step$\ 2'.$}] For $a=1$, obtain a bootstrap replicate
of $\hat{q}_{1,\xi\mid A=1}(\hat{\theta})$, $\hat{q}_{1,\xi\mid A=1}^{*}(\hat{\theta}^{*})$,
by solving
\[
\hat{F}_{1}^{*}(q\mid A=1)=n_{1}^{-1}\sum_{i=1}^{n}\omega_{i}^{*}A_{i}\bone(Y_{i}\leq q)=\xi.
\]
For $a=0$, obtain a bootstrap replicate of $\hat{q}_{0,\xi\mid A=1,\dsm}(\hat{\theta})$,
$\hat{q}_{0,\xi\mid A=1,\dsm}^{*}(\hat{\theta}^{*})$, by solving
\begin{multline*}
\hat{F}_{0,\dsm}^{*}(q\mid A=1)=n_{1}^{-1}\sum_{i=1}^{n}\omega_{i}^{*}A_{i}\hat{F}_{0}\{q;S_{i}(\hat{\theta}^{*})\}\\
+n_{1}^{-1}\sum_{i=1}^{n}\omega_{i}^{*}\bone(A_{i}=0)M^{-1}K_{S(\hat{\theta}),i}\left[\bone(Y_{i}\leq q)-\hat{F}_{0}\{q;S_{i}(\hat{\theta}^{*})\}\right]=\xi,
\end{multline*}
for $q$. Then a bootstrap replicate of $\hat{\Delta}_{\ATT,\xi,\dsm}(\hat{\theta})$
is $\hat{\Delta}_{\ATT,\xi,\dsm}^{*}(\hat{\theta}^{*})=\hat{q}_{1,\xi\mid A=1}^{*}(\hat{\theta}^{*})-\hat{q}_{0,\xi\mid A=1,\dsm}^{*}(\hat{\theta}^{*})$.
\end{description}

\section{Figure\label{sec:Figure}}

\begin{figure}
\caption{\label{fig2:lalonde}Histogram of outcome variable (re78)}

\centering{}\includegraphics{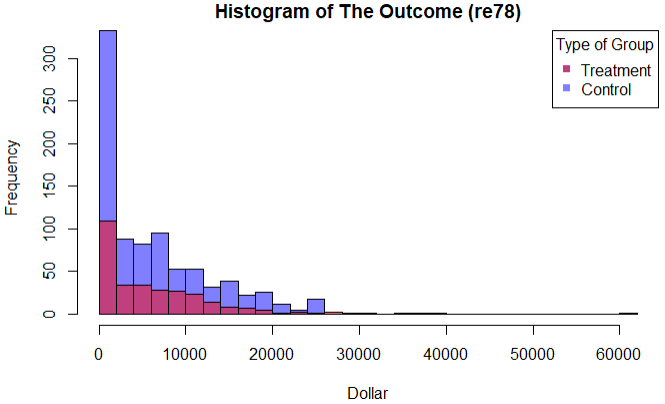}
\end{figure}

Figure \ref{fig2:lalonde} shows that the outcome distributions are
highly skewed in the data from the job training program. 

\end{document}